\documentclass[12pt]{article}

\usepackage[letterpaper,margin=3cm]{geometry}

\usepackage{authblk}

\usepackage[parfill]{parskip}
\usepackage{setspace}
\usepackage{soul}

\usepackage{booktabs}
\usepackage{multirow}
\usepackage{geometry}
\usepackage{sectsty}
\sectionfont{\fontsize{16}{22}\selectfont}

\usepackage[labelfont=bf,font={small,stretch=0.95}]{caption}

\usepackage[super,sort,compress,numbers]{natbib}

\makeatletter
\renewcommand\@biblabel[1]{#1.}
\makeatother

\usepackage{array}

\usepackage{gensymb}
\usepackage{amsmath}
\usepackage{amssymb}
\usepackage{siunitx}
\usepackage{mathtools}
\usepackage{comment}
\usepackage{hyperref}
\usepackage{float}

\usepackage{graphicx}
\usepackage{multirow}

\usepackage[version=4]{mhchem}

\usepackage{mathptmx}

\usepackage{xcolor}
\usepackage{soul}
\usepackage{booktabs}

\begin{document}

\title{\fontsize{20}{26}\selectfont\bf Data-driven atomistic modelling of hybrid halide perovskite passivation}

\author[1]{Laura-Bianca Pa\c{s}ca}
\author[2]{Henry J. Snaith}
\author[1]{Volker L. Deringer\thanks{E-mail: \url{volker.deringer@chem.ox.ac.uk}}}

\affil[1]{Inorganic Chemistry Laboratory, Department of Chemistry, University of Oxford, \protect\\ Oxford, UK}
\affil[2]{Department of Physics, University of Oxford, Oxford, UK}

\date{}

\maketitle

\setstretch{1.2}\selectfont

{\bf 
Molecular passivation of surface defects is key to improving the optoelectronic performance of hybrid halide perovskite materials, but the underlying atomistic mechanisms are incompletely understood. While machine-learned interatomic potentials are now widely used to simulate complex molecular and crystalline systems, their application to experimentally-realistic scenarios -- such as molecules coordinating to perovskite surfaces -- is still far from trivial. Here, we describe a multistep training pipeline, resembling continuous fine-tuning used for large language models, to underpin atomistic modelling and computational experiments in this domain. Our protocol involves two components: (i) a large, curated, and open dataset of diverse metal and hybrid halide perovskite structures (`hyP-26'); and (ii) a small, specialised dataset for an amino-silane molecule passivating the surface, providing highly specific information for fine-tuning. We apply this approach to explore collective behaviour at a mixed-composition halide perovskite surface passivated with a varying coverage of amino-silane molecules, revealing an evolution of interactions with increasing molecular surface coverage.
}

\clearpage
\setstretch{1.5}
\section*{Introduction}

Hybrid perovskite materials are at the forefront of emerging solar-cell research. Despite their favour\-able optoelectronic properties, a well-known challenge is their decomposition under humidity, heat, prolonged illumination, electrical bias, and especially combinations of these conditions.\cite{helalmiah_key_2025} To improve operational durability, a central experimental strategy for reducing ion migration and degradation is through molecular passivation of interfaces, whereby passivating molecules bind to surface defects.\cite{jariwala_reducing_2021} Amino-silanes have proven effective in this role,\cite{bandgap_science} with (3-aminopropyl)trimethoxysilane (AEAPTMS) in particular having been shown to improve the optoelectronic properties of perovskite films following surface treatment, and considerable improvement in long-term stability.\cite{huang_passivation_2025}

From a computational chemistry perspective, molecular passivation presents a substantial challenge: it requires the modelling of bulk disorder, surface reconstructions, and chemically distinct adsorbates, using large simulation cells to capture the full range of relevant structures. Large-scale simulations are now routinely accessible in principle using machine-learned interatomic potentials (MLIPs);\cite{Lu_pflops_2021, morrow_defects_2024} an overview of MLIPs for hybrid halide perovskites is given in Ref.~\citenum{Bian2026}. Fine-tuning a general-purpose pre-trained or `foundational' model (FM) on a small, representative dataset can yield an accurate system-specific potential at a fraction of the cost of directly training a material-specific MLIP.\cite{deng_overcoming_2024, batatia_foundation_2025, aisnada_cost-effective_2025, Tompa2026} Despite their broad success, current FMs struggle in chemically heterogeneous, disordered, or interfacial environments\cite{echeverri_restrepo_applicability_2025, kempen_how_2025} -- precisely the type of conditions encountered in molecular passivation of (poly)crystalline surfaces. To deal with this challenge, bespoke datasets and training approaches have been explored: for example, for heterogeneous catalysis applications, \cite{Wu_CLAM_2025} and indeed recently for halide perovskites.\cite{Biswas_2026} Material-specific datasets can supply the structural motifs that are sparsely represented in large pre-training datasets, such as specific surface terminations or defect configurations of interest.

In large language model (LLM) research, the continual training of models through a protocol called continual fine-tuning (CFT) has been proposed for adapting to ever-expanding and evolving datasets without retraining the model from scratch.\cite{zhang_citb_2023, aggarwal_exploring_2024, guan_multi-stage_2025} Similarly, constructing ML models with varying levels of specialisation for a target task can be achieved through a CFT-inspired approach. Such strategies have already been explored for chemistry applications in the development of an LLM specialised in materials research tasks;\cite{ahlawat_family_2026} for training MLIPs, the concept of `lifelong learning' protocols for integrating new data has been explored for chemical reaction networks. \cite{eckhoff_lifelong} Strategies for preventing `catastrophic forgetting', the loss of knowledge gained in previous stages of training, have recently been proposed for continual learning in MLIPs,\cite{kim_efficient_2025, liu_ai-ready_2026} and repeated, iterative fine-tuning has been studied. \cite{Wong2026} However, questions remain: is there a role for wider-ranging, yet material-specific datasets in fine-tuning protocols, whereby combining datasets with different levels of specialisation can lead to improved performance on the target system for MLIP modelling? And would such an approach still be required as the size and diversity of pre-training datasets increases, up to a point where `zero-shot' performance already approaches a plateau for general applications?

Herein, we describe a machine-learning-based modelling framework that embeds material-specific datasets within a fine-tuning protocol for modelling complex, application-driven systems -- here, hybrid perovskites. Our approach has two stages:
(i) the construction of a diverse, curated dataset (which we call `hyP-26'), containing representative metal and mixed-composition (mixed cation and halide) hybrid perovskite structures, including various disordered phases; and 
(ii) the application of `hyP-26' in a CFT-like training approach to produce an MLIP specialised for AEAPTMS passivation at a mixed hybrid perovskite surface. 
We demonstrate how this protocol can be applied to simulate the binding modes of AEAPTMS molecules with varying surface coverages. Our simulations allow us to model the collective, dynamic behaviour of multiple amino-silane molecules binding to and disrupting the crystalline hybrid perovskite lattice.
More generally, we provide a blueprint for ML-driven simulations of complex, experimentally relevant systems.

\section*{A dataset for modelling hybrid perovskite passivation}

Many large, open datasets of chemical structures are now available and used to pre-train MLIPs for general chemistry applications.\cite{oc20_2021, deng_2023_chgnet, barroso-luque_open_2024, omol25_2025, cavignac_ai-driven_2026} Starting from such pre-trained FMs based on the MACE architecture,\cite{Batatia_mace} we employed different fitting approaches to systematically study the role of the datasets used at each stage of the fine-tuning protocol (the latter being described in detail in Supplementary Note 1).

\begin{figure*}[t]
\centering
  \includegraphics[width=\textwidth]{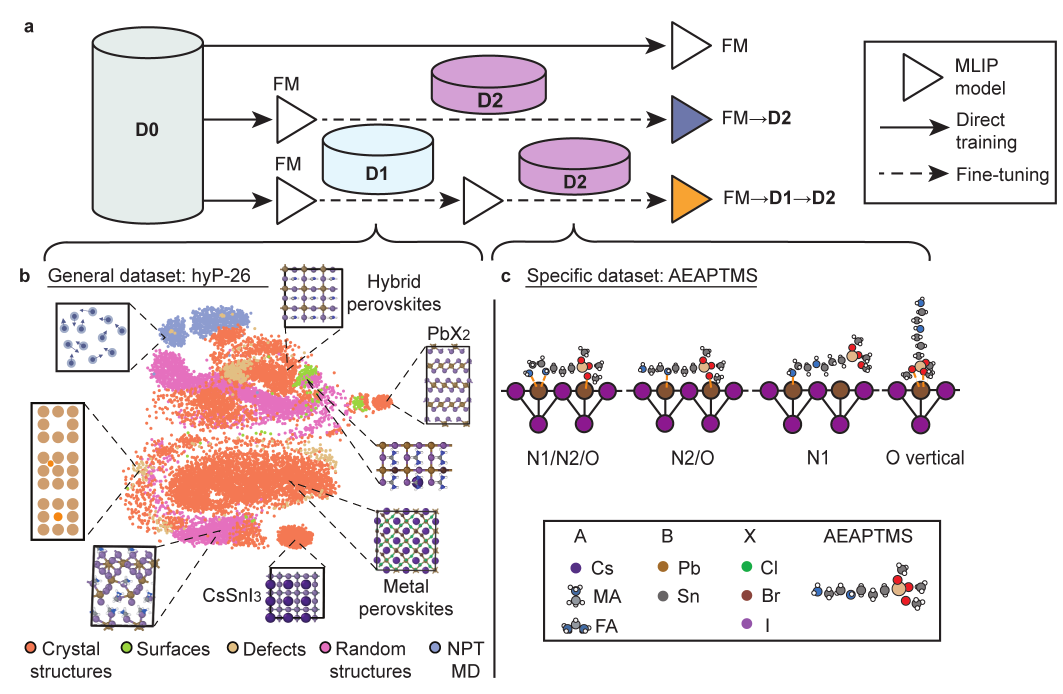}
  \caption{\textbf{Fine-tuning pipelines for ML-driven simulations of hybrid perovskite passivation}. 
  (\textbf{a}) Schematic showing the different fine-tuning protocols compared in the present work: the zero-shot foundation model (`FM'), the fine-tuned model (`FM$\rightarrow$\textbf{D2}'), and the continuously fine-tuned model (`FM$\rightarrow$\textbf{D1}$\rightarrow$\textbf{D2}'). 
  (\textbf{b}) The composition of the `hyP-26' dataset, here used as the material-specific dataset, \textbf{D1}. The two-dimensional representation of the configurational space is constructed with the UMAP algorithm \cite{mcinnes_umap_2018} based on the Smooth Overlap of Atomic Positions (SOAP) similarity metric,\cite{bartok_soap} showing the clustering of structurally-similar systems included in the dataset. 
  (\textbf{c}) The composition of the specialised dataset, \textbf{D2}, focused on hybrid perovskite surface passivation with AEAPTMS. Structural sketches show the main coordination types studied (denoted, from left to right: `N1/N2/O', `N2/O', `N1', and `O vertical').}
  \label{fgr:overview}
\end{figure*}

A schematic illustration of the three main protocols is shown in Fig.~\ref{fgr:overview}a. We denote the extensive pre-training dataset \textbf{D0} in this scheme. For simulating known materials that are well-represented in \textbf{D0}, pre-trained MLIPs can sometimes be used out-of-the-box (`zero-shot'), without further fine-tuning. For more complex systems or target properties, a small subset of specialised data, denoted here as \textbf{D2}, is now predominantly used to fine-tune the FM (`FM$\rightarrow$\textbf{D2}'; dark blue in Fig.~\ref{fgr:overview}a). This approach replaces the use of a curated, material-specific dataset, \textbf{D1}, which would previously have been laboriously assembled and used to train an MLIP from scratch. Herein, we also explore a stepwise strategy: to first adapt the FM to the intermediate dataset \textbf{D1}, and then subsequently fine-tune it on \textbf{D2} in a CFT-like scheme (`FM$\rightarrow$\textbf{D1}$\rightarrow$\textbf{D2}'; orange in Fig.~\ref{fgr:overview}a). As a point of reference, we furthermore fit a model from scratch, directly for the domain of interest (that is, on both \textbf{D1} and \textbf{D2}). 

We stress that our study focuses on the model's performance on the specialised application targeted by \textbf{D2}, rather than aiming to preserve the same level of performance on the more general application represented by \textbf{D1}. For such a use case, `naive' fine-tuning, as also employed in the present work, has been found to perform well without requiring multi-head replay or other strategies for retaining generalisability beyond the target application.\cite{Tompa2026} Our work now interrogates whether the CFT loop adds any benefit compared to direct FT practices in the limit of pre-training datasets with extensive coverage of the relevant configurational space.

\setstretch{1.5}

To investigate the effect of the general pre-training dataset, \textbf{D0}, we applied the training protocol to different MACE-based FMs. Specifically, we considered \texttt{MACE-MP-0b3},\cite{batatia_foundation_2025} pre-trained on the \texttt{MPTrj} dataset,\cite{deng_2023_chgnet} and \texttt{MACE-MH-1},\cite{batatia_cross_2025} trained using a multi-head fine-tuning strategy on a pre-training dataset with a much more diverse configurational space. Two additional models, \texttt{MACE-MPA-0} and \texttt{MACE-OMAT-0}, are discussed in the Supplementary Information. 

The \texttt{MPTrj} dataset consists of $\approx$~1.6M structures sampled from relaxation trajectories of Materials Project entries.\cite{deng_2023_chgnet} While this dataset has been extensively used for pre-training MLIPs, its construction emphasises near-equilibrium inorganic crystalline configurations and includes a small fraction of organic environments (with only 1.8\% of the dataset composition consisting of organic molecular crystals \cite{taniguchi_knowledge_2025}). As a result, it underrepresents chemical environments relevant to heterogeneous or molecular–inorganic systems, such as those considered here. 
By contrast, the datasets underlying \texttt{MACE-MH-1} \cite{batatia_cross_2025} cover a broader configurational space, including inorganic crystals, surface structures, organic molecules and reactive organic systems,\cite{oc20_2021, barroso-luque_open_2024, omol25_2025, kaplan_foundational_2025, cavignac_ai-driven_2026} and it has been shown to achieve an improved out-of-the-box performance on these systems compared to models pre-trained on datasets with a smaller configurational coverage.\cite{batatia_cross_2025} The surface systems and organic molecules each comprise between 10--15\% of the dataset used in the multi-head fine-tuning stage of the FM generation protocol.\cite{batatia_cross_2025} A previous study proposed a framework for combining the fine-tuning \texttt{MACE-MH-1} with delta-learning for predicting kinetic and thermochemical parameters of reactions in heterogeneous catalysis,\cite{johnson_fusing_2026} another domain requiring accurate modelling of surfaces, and both organic and inorganic chemical components.

By comparing the FMs pre-trained on these two datasets within the same CFT-like framework, we assess how the degree of coverage of \textbf{D0} influences the effectiveness of subsequent fine-tuning. In particular, this allows us to contrast the case when \textbf{D0} underrepresents the chemical domain of interest with that in which the pre-training data already covers a highly-diverse chemical space, providing insight into when CFT on the domain-specific dataset \textbf{D1} offers tangible advantages over conventional fine-tuning strategies.

To this end, we constructed a general hybrid perovskite-specific dataset, introduced here as `hyP-26', which represents the \textbf{D1} training data required for the fitting protocol. As illustrated in Fig.~\ref{fgr:overview}b, `hyP-26' contains mixed-composition hybrid perovskite structures, the binary crystalline phases of the lead halides (PbX$_{2}$), higher-temperature molecular dynamics (MD) structures, defect and surface systems, as well as random structures, with the aim of covering chemical environments possibly present in disordered phases or interfacial systems. We show the distribution of the structures in a two-dimensional representation of the SOAP similarity metric, as a per-structure average:\cite{bartok_soap} this means that related structures will cluster together in the colour-coded map. The same representation colour-coded with respect to the cohesive energy of each structure is shown in Fig.~S1. The generation of the dataset is described in detail in Supplementary Note 2 and presented in Tables~S1--S4.

The specific dataset \textbf{D2}, illustrated in Fig.~\ref{fgr:overview}c, corresponding to the downstream task of modelling hybrid perovskite surface passivation, was generated using a combination of iterative MD training and sampling of relaxation trajectories (Supplementary Note 2). The passivating molecule considered here is AEAPTMS, which has been reported as an effective amino-silane passivator for hybrid perovskite surfaces.\cite{bandgap_science} Most studies of amino-silane passivation were conducted on a mixed hybrid perovskite composition \cite{Shi_improves, bandgap_science} investigated for its use in the production of tandem solar cells;\cite{mcmeekin_crystallization_2017} we therefore employed a similar composition of Cs$_{0.12}$FA$_{0.88}$PbI$_{0.75}$Br$_{0.25}$. Starting with an AEAPTMS molecule placed on different sites at the surface, the system is relaxed into four representative coordination environments, illustrated in Fig.~\ref{fgr:overview}c: bidentate coordination involving two nitrogen atoms and an oxygen atom binding to neighbouring Pb sites (`N1/N2/O'); nitrogen–oxygen coordination at neighbouring Pb sites (`N2/O'); monodentate coordination via the first nitrogen atom (`N1'); and vertical coordination through oxygen atoms (`O vertical'). We assume that the multidentate binding configurations are favourable for stable passivation since the molecules are less likely to entirely desorb from the surface over time.

We found that including snapshots from relaxation trajectories in the training dataset was necessary for the \texttt{MACE-MP-0b3}-derived models' ability to identify the `N1/N2/O' configuration type during MLIP-driven MD or structural relaxations (discussed below). The 'direct-hyP-26+AEAPTMS' and \texttt{MACE-MH-1}-derived models are able to identify this binding mode without its explicit inclusion in the training dataset; a comparison of the models' performance in predicting the single-point energies of the different configuration types before and after the inclusion of the relaxation trajectory snapshots is provided in Fig.~S2: the \texttt{MACE-MP-0b3}-derived models overpredict the energies by over 10 meV/atom without the inclusion of the relaxation snapshots, whereas the \texttt{MACE-MH-1}-based models already show a sub-10 meV/atom energy prediction compared to DFT without the explicit inclusion of the relaxed configuration types. 

Our \textbf{D2} dataset captures the complex coordination environments associated with molecular adsorption at hybrid perovskite surfaces, constituting a heterogeneous, multicomponent test case that lies beyond the scope of typical pre-training datasets. 

\section*{Validation across general and specific tasks}

As a first step in validating the different training protocols, we assess the root mean square errors (RMSEs) of the force components predicted by the potentials compared to DFT values. The energy and force RMSEs for the different stages of the fine-tuning protocol are reported in Tables~S5 and S6. We assess the retention of accuracy on more general tasks for hybrid perovskite modelling, and compare this with the accuracy on the specific target task of molecular passivation. To this end, we select structures from the \textbf{D1} (`hyP-26') test set with chemical environments relevant to the specialised task, namely perovskite surface structures and \ce{MAPbI3} defect configurations. Surface and defective environments are sampled in the \textbf{D2} dataset as part of the surface--adsorbate structures, but the individual constitutents, viz.\ isolated surfaces and isolated molecules, are not. As such, testing on these related general tasks provides an informative test of transferability. 

\begin{figure*}[t]
\centering
  \includegraphics[width=\textwidth]{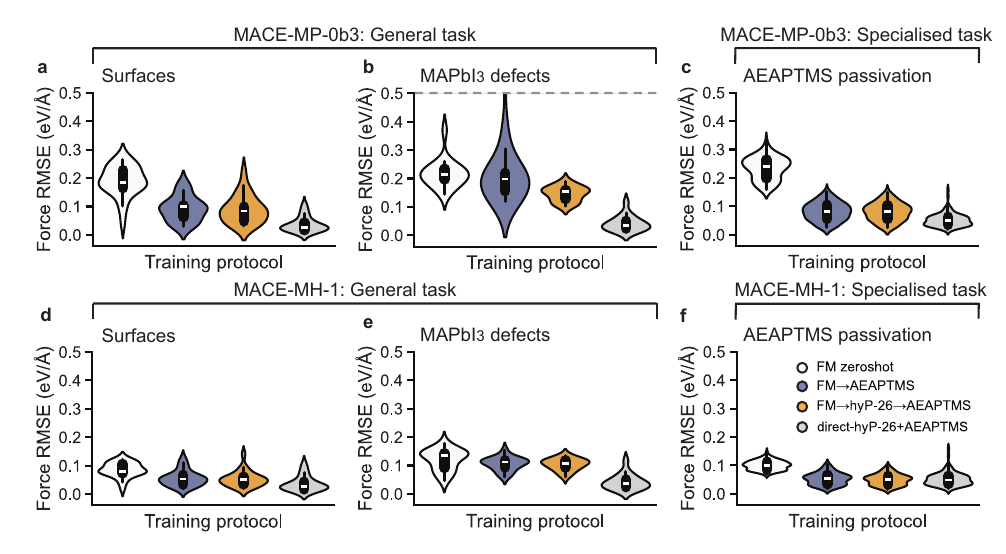}
 \caption{\textbf{Influence of fitting protocol on model performance for general and specialised tasks.} Violin plots show errors expressed as force RMSEs for surface structures and \ce{MAPbI3} defect systems (`hyP-26' test set), as well as the \textbf{D2} specialised test set. Panels (\textbf{a})--(\textbf{c}) show results for \texttt{MACE-MP-0b3} as the starting FM, while panels (\textbf{d})--(\textbf{f}) show results for \texttt{MACE-MH-1}. The dashed grey line in panel (\textbf{b}) marks the presence of values outside the $y$-axis range shown, up to 0.72 eV/\AA{}.}
  \label{fgr:RMSE}
\end{figure*} 

When using \texttt{MACE-MP-0b3} as the starting point in our training protocol, we observe significant variations in model performance between the different training approaches (Fig.~\ref{fgr:RMSE}a--c). Without fine-tuning, the zero-shot FM shows similar median force RMSEs for the hybrid perovskite surfaces and \ce{MAPbI3} defect systems ($\approx$~200 meV/\AA{}), with an increase in force RMSE for the specialised task (240 meV/\AA{}). Fine-tuning \texttt{MACE-MP-0b3} using the `FM$\rightarrow$ AEAPTMS' protocol decreases the error on the specialised task by $\approx$~150 meV/\AA{}, with good transferability to the surface systems. However, the errors show little improvement in the case of the \ce{MAPbI3} defect systems. Switching to the CFT protocol (`FM$\rightarrow$hyP-26$\rightarrow$AEAPTMS') improves the performance on the \ce{MAPbI3} defect structures and significantly reduces the variance in the force predictions for this system. Direct fitting to the combined `hyP-26' and specialised \textbf{D2} datasets yields the lowest RMSE across all three applications. As this approach does not involve sequential training stages, it avoids catastrophic forgetting of previously-learned information. However, training from scratch requires extensive, carefully curated datasets and long training times, typically incurring much higher computational cost compared to fine-tuning approaches. Moreover, the resulting potential shows reduced transferability to chemically-related amino-silane passivators in the specialised task (Fig.~S4).

By contrast, when using \texttt{MACE-MH-1} as the starting point, the numerical differences between the directly-trained and fine-tuned potentials are substantially reduced across all test cases (Fig.~\ref{fgr:RMSE}d–f), consistent with the lower zero-shot RMSEs of this model, which are two to three times lower than those of the zero-shot \texttt{MACE-MP-0b3} for the same systems. The variance is reduced across all three fitting approaches that depend on the choice of starting FM. In this case, the additional benefits of CFT over conventional fine-tuning for surface and defect environments are limited, suggesting that the wider coverage of the pre-training data underpinning \texttt{MACE-MH-1} provides good transferable accuracy as measured by error metrics.

\section*{Energetics of passivator--surface interactions}

To study the energy prediction performance of the different MLIPs, we chose a metric relevant to application: the predicted adsorptions energies, $E_\textnormal{ads}$,  for representative AEAPTMS coordination types at the perovskite surface.
The assessment proceeds in two stages of increasing difficulty: (i)~static single-point energy evaluation of DFT-relaxed structures; and (ii)~MLIP-driven structural relaxation referenced against DFT ground-truth configurations.

In the first stage, we compare MLIP single-point adsorption energies evaluated on DFT-relaxed structures to DFT reference values. The adsorption energy is defined as
\begin{equation}
    E_{\textnormal{ads}} = E_{\textnormal{system}} - (E_{\textnormal{surface}} + E_{\textnormal{molecule}}) 
\end{equation}
where the three terms correspond to the relaxed adsorbate–surface system, the clean surface, and the isolated molecule, respectively.

\begin{figure*}[p]
\centering
  \includegraphics[width=\textwidth]{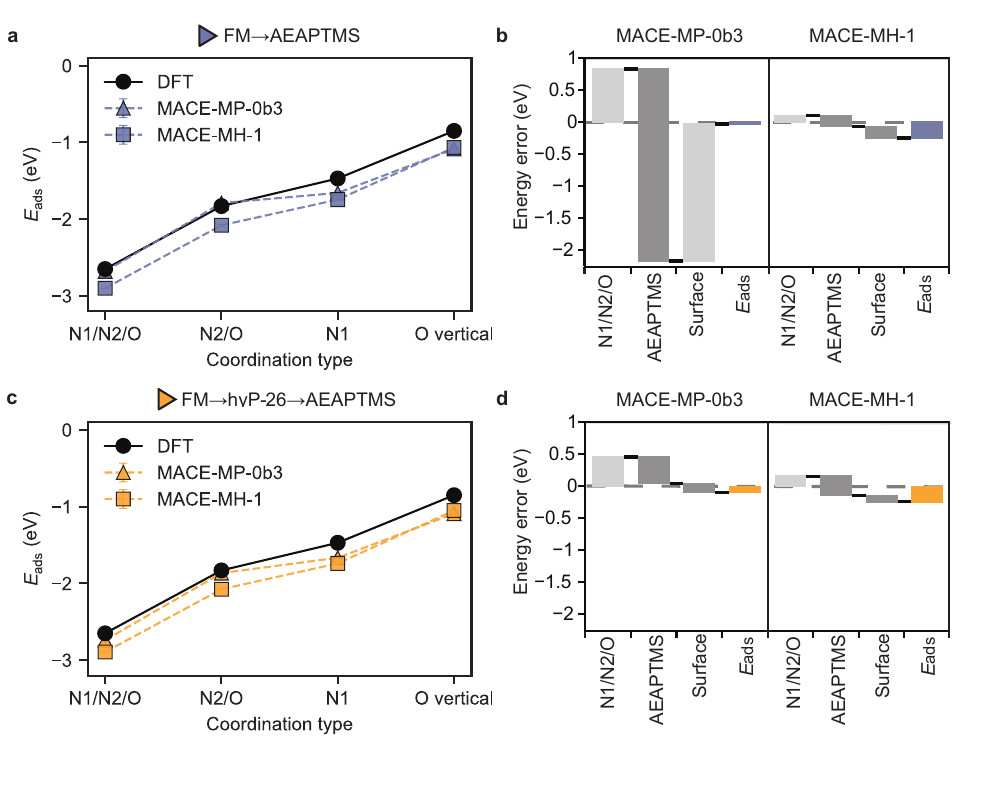}
  \caption{\textbf{Error cancellation in MLIP static adsorption energy predictions.} 
  (\textbf{a}) Adsorption energies predicted for the different coordination modes, with MLIP-predicted energies for the DFT-relaxed configurations, using the `FM$\rightarrow$AEAPTMS' fitting protocol. (\textbf{b}) Waterfall plot breaking down the energy errors contributing to the MLIPs’ deviation from DFT-predicted $E_{\textnormal{ads}}$ for the `N1/N2/O' system. The light grey shading shows a positive error compared to the DFT value (energy overprediction), while the darker grey shading shows a negative error (energy underprediction). (\textbf{c})--(\textbf{d}) As above, but now for the `FM$\rightarrow$hyP-26$\rightarrow$AEAPTMS' fitting protocol.}
  \label{fgr:AEAPTMS-E}
\end{figure*} 

The $E_\textnormal{ads}$ prediction of the `FM$\rightarrow$AEAPTMS' models is shown in Fig.~\ref{fgr:AEAPTMS-E}a. The protocol using \texttt{MACE-MH-1} as the starting FM shows a trend consistent with DFT for the relative stability of the different coordination types, with an approximately uniform overprediction of the binding stability by about --0.25 eV. In the case of \texttt{MACE-MP-0b3}, the `N1'-coordinating system's binding stability is overpredicted relative to the `N1/N2/O' and `N2/O' systems, but the latter show values nearly identical to the DFT $E_\textnormal{ads}$ predictions.

However, a more detailed analysis reveals that the close alignment with DFT values for the \texttt{MACE-MP-0b3} `FM$\rightarrow$AEAPTMS' model arises from a substantial cancellation of errors in the calculation of $E_{\textnormal{ads}}$, as revealed by the waterfall chart in Fig.~\ref{fgr:AEAPTMS-E}b. The per-atom energy errors, more routinely reported in meV/atom, are shown in the Supplementary Information (Table~S10). The isolated molecule and surface slab present the largest errors, which are significantly reduced when starting the training from the \texttt{MACE-MH-1} FM, pre-trained on a dataset which includes organic molecules and surface systems.

The $E_\textnormal{ads}$ predictions of the `FM$\rightarrow$hyP-26$\rightarrow$AEAPTMS' models, presented in Fig.~\ref{fgr:AEAPTMS-E}c, show a very similar trend to the conventional fine-tuning protocol characterised in Fig.~\ref{fgr:AEAPTMS-E}a. Yet, the analysis of energy errors illustrated by the waterfall chart (Fig.~\ref{fgr:AEAPTMS-E}d) shows a significant improvement compared to `FM$\rightarrow$AEAPTMS' when starting from the \texttt{MACE-MP-0b3} FM, which is masked by the cancellation of errors. In the case of \texttt{MACE-MH-1}, only the energy error on the surface system shows an improvement when adopting the CFT protocol.

A similar analysis for the `direct-hyP-26+AEAPTMS' model, and additionally for models derived from the \texttt{MACE-MPA-0} and \texttt{MACE-OMAT-0} FMs, are provided in the Supplementary Information (Fig.~S5 and Fig.~S6 and Tables~S7--S9).
The zero-shot model results are not presented here due to the large deviations from the DFT reference, with adsorption energy errors exceeding 2 eV when using \texttt{MACE-MP-0b3} in the fine-tuning protocol, but it is also available in the Supplementary Information (Fig.~S5).

\begin{figure*}[t]
\centering
  \includegraphics[width=\textwidth]{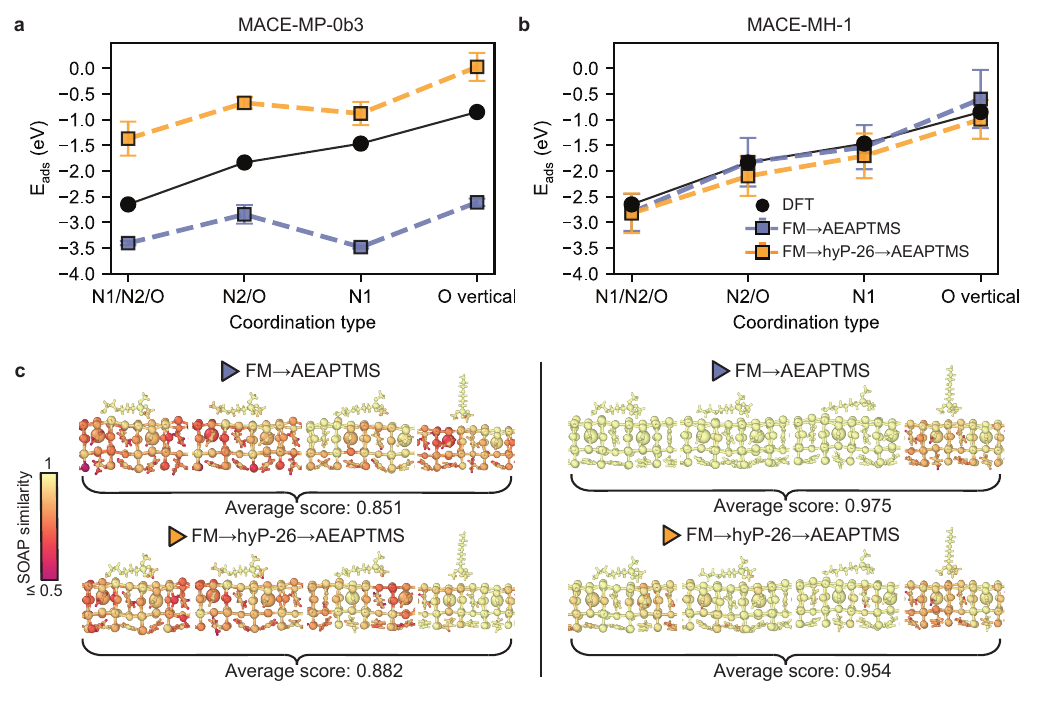}
  \caption{\textbf{MLIP-driven relaxation of the amino-silane molecule passivating the hybrid perovskite surface.} (\textbf{a}) Adsorption energies predicted for the different coordination modes, with \texttt{MACE-MP-0b3}-derived MLIP-predicted energies for the MLIP-relaxed configurations. The DFT-predicted energies of the DFT-relaxed structures are shown in black. (\textbf{b}) Same, but for \texttt{MACE-MH-1}-derived MLIPs. (\textbf{c}) Structural similarity of the MLIP-relaxed structures compared to the DFT-relaxed systems for each coordination type. Atom colors indicate their SOAP similarity to the corresponding atoms in the ground-truth DFT-relaxed configuration, as visualized using OVITO.\cite{ovito}}
  \label{fgr:SOAP}
\end{figure*}

In the second stage of evaluation, we use the MLIPs to drive structural relaxation into the four different coordination types. This scenario challenges the MLIP's ability to accurately map the potential energy surface across the relaxation pathway: a failure to do so will cause the MLIP to converge to a geometry inconsistent with that obtained by DFT relaxation. The values of $E_{\textnormal{ads}}$ predicted by the MLIP for the MLIP-relaxed structures are compared to the DFT-predicted energies of the DFT-relaxed structures in Figs.~\ref{fgr:SOAP}a--b. (The DFT single-point energies for the same MLIP-relaxed structures show a similar trend: see Figs.~S7--S8.) 

For \texttt{MACE-MP-0b3}-derived models (Fig.~\ref{fgr:SOAP}a), the fine-tuned MLIPs relax the system into configurations with $E_{\textnormal{ads}}$ deviations of up to 1 eV from the DFT reference. The `FM$\rightarrow$AEAPTMS' model shows the largest energetic error for the surface system (Table~S10), suggesting that the MLIP has failed to find the energy minimum during relaxation. The cancellation of errors of similar magnitudes for the `N1/N2/O' and `N2/O' systems and the surface structure lead to the perceived agreement between the DFT- and MLIP-predicted $E_{\textnormal{ads}}$ value for these configuration types. Adopting the CFT-like protocol (Fig.~\ref{fgr:overview}a) significantly reduces the error on the surface structure, with less improvement shown for the surface--adsorbate system. 

When using \texttt{MACE-MH-1} as the starting point, these errors are substantially reduced, with maximum $E_\textnormal{ads}$ deviations below 0.5 eV (Fig.~\ref{fgr:SOAP}b). This improvement relative to \texttt{MACE-MP-0b3}-derived models, as well as to the directly-trained potential (Fig.~S7) is attributed to the inclusion of reactive organic systems, surface slabs, and adsorbate complexes in the pre-training data of \texttt{MACE-MH-1}. \cite{batatia_cross_2025} Zero-shot results are reported in Fig.~S9.

However, such high numerical accuracy may not always be required when performing qualitative chemical analyses. FMs or fine-tuned models are now commonly used to obtain broad structural insights and trends in varied systems of interest.\cite{echeverri_restrepo_applicability_2025, ng_accelerated_2025, kempen_how_2025} To decouple structural validation from the validation of predicted energies, we therefore quantify the similarity between MLIP-relaxed and DFT-relaxed structures using the SOAP similarity metric, \cite{bartok_soap} as shown in Fig.~\ref{fgr:SOAP}c. In this metric, a score of 1 corresponds to identical local environments, while a score of 0 indicates completely dissimilar environments. The atoms are colour-coded according to their SOAP similarity to the corresponding atom in the respective DFT-relaxed system. Across all models, the largest deviations from the DFT structures are observed within the hybrid perovskite surface slab. Consistent with the $E_{\textnormal{ads}}$ analysis, when starting from \texttt{MACE-MP-0b3}, the CFT MLIP produces structures more closely approaching the DFT references than those obtained using conventional FT on \textbf{D2} alone. When derived from \texttt{MACE-MH-1}, both fine-tuned models show a significantly improved performance in the structural relaxation task compared to the \texttt{MACE-MP-0b3}-based models; moreover, they surpass the performance of the directly-trained model (Fig.~S10). The `FM$\rightarrow$AEAPTMS' MLIP in fact averages a slightly higher SOAP similarity score over the four different coordination types compared to the CFT protocol (`FM$\rightarrow$hyP-26$\rightarrow$AEAPTMS'). We link this observation to the already adequate performance of the zero-shot model: interestingly, used out-of-the-box, the performance of \texttt{MACE-MH-1} in structural relaxation matches that of \texttt{MACE-MP-0b3} fine-tuned for AEAPTMS (Fig.~S11). In contrast, the zero-shot \texttt{MACE-MP-0b3} model produces pronounced structural distortions, including Pb--I bonds elongated by up to $\approx$ 30\% relative to the DFT result, particularly along the surface-normal direction.
The average similarity scores of the configurations obtained with potentials based on different FMs are given in Table S12.

\section*{Data-driven atomistic modelling of molecular passivation}

To show its application in a real-world chemical scenario, we use the \texttt{MACE-MH-1}-derived 'FM$\rightarrow$hyP-26$\rightarrow$AEAPTMS' model to run MD simulations of a mixed hybrid perovskite surface passivated with an increasing number of AEAPTMS molecules; this simulation set-up resembles an experimental increase in deposition time and surface coverage. For each chosen surface passivation coverage, 5 independent MLIP-driven MD runs were performed holding at 450~K for 500~ps, followed by structural relaxation with the MLIP. As such, the final structures reflect the transient configurations adopted by AEAPTMS as it diffuses and reorients on the surface at the finite simulation temperature, rather than representing the global energy minimum alone. For the purposes of this study, we have focused on a Pb-terminated Cs$_{0.12}$FA$_{0.88}$PbI$_{0.75}$Br$_{0.25}$ surface without additional vacancies or interstitial defects; studying the effect of such defective surfaces would be a valuable addition for future work. 

\begin{figure*}[p]
\centering
  \includegraphics[width=\textwidth]{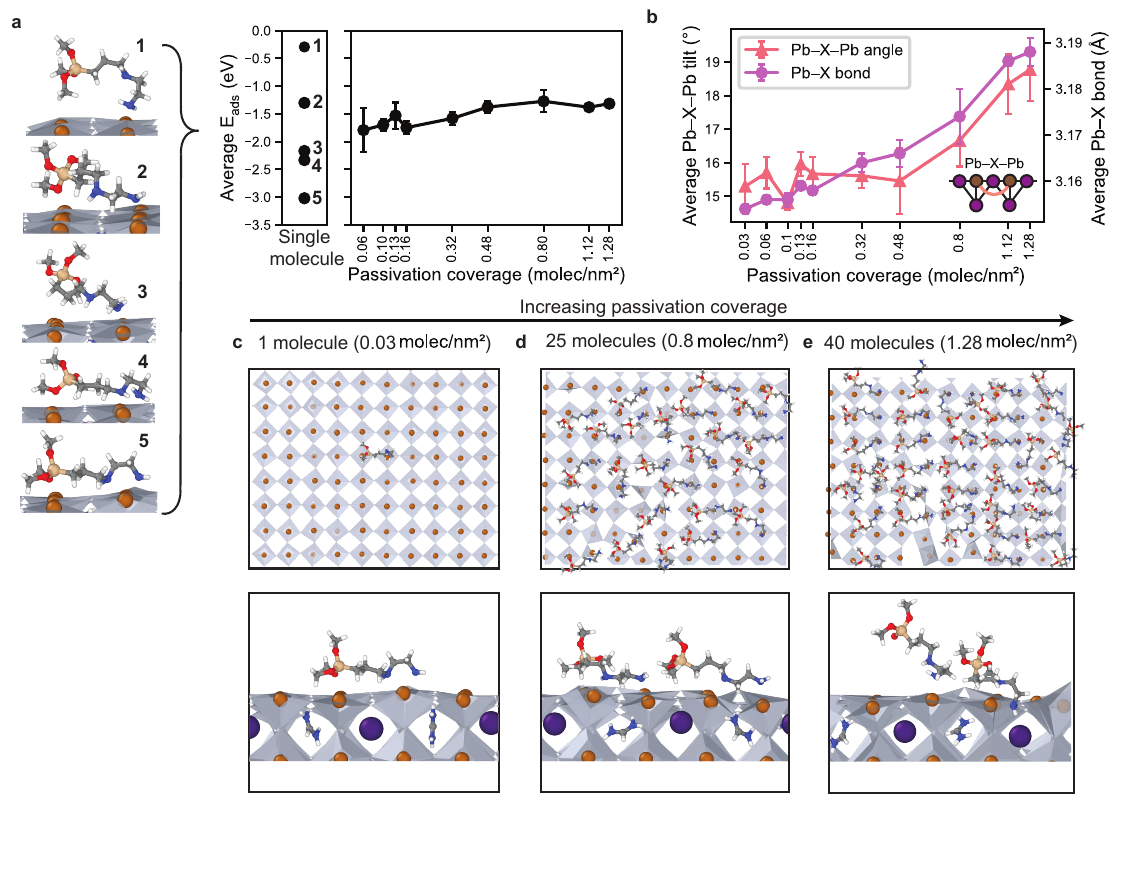}
  \caption{\textbf{The link between increased passivation coverage and structural disruption of the perovskite surface.} (\textbf{a}) Adsorption energy as a function of passivation coverage, averaged over 5 independent MD runs and subsequently relaxed using the \texttt{MACE-MH-1}-derived `FM$\rightarrow$hyP-26$\rightarrow$AEAPTMS' model. The left panel shows the diversity in observed coordination types for a single molecule passivating the surface. (\textbf{b}) Change in Pb--X--Pb angles (octahedral tilts) and Pb--X bond lengths at the hybrid perovskite surface with increasing passivation coverage. (\textbf{c}) (Top panel) Top view of the MLIP-driven MD simulation with a single molecule coordinating at the surface. (Bottom panel) Close-up snapshot from the simulation, showing the molecule coordinating at the surface. (\textbf{d}) Same, but for a passivation coverage of 0.80~molecules/nm$^{2}$. (\textbf{e}) Same, but for a passivation coverage of 1.28~molecules/nm$^{2}$, where the molecule starts penetrating into the surface. Note that the lower panels show only selected molecules; the overall surface coverage can be seen in the top-view panels above. Structural visualisations were produced using OVITO. \cite{ovito}}
  \label{fgr:application}
\end{figure*}

The variation of $E_{\textnormal{ads}}$ with increasing passivation coverage is shown in Fig.~\ref{fgr:application}a, this time calculated by dividing the expression in Equation 1 by the number of molecules present at the surface. For a single molecule, the five independent runs yield a range of coordination geometries (shown on the left; \textbf{1}--\textbf{5}), resulting in the substantial variance of $E_{\textnormal{ads}}$. The multidentate `N1/N2/O' coordination (\textbf{5}) shows an $E_{\textnormal{ads}}$ value of just under $-3$ eV, consistent with previously reported ab initio results for AEAPTMS binding to a pristine surface.\cite{bandgap_science} As the number of passivating molecules increases, contributions from different coordination types average out, the variance decreases, and $E_{\textnormal{ads}}$ stabilises at $\approx -1.5$ eV. 

Notably, increasing AEAPTMS surface coverage exerts a measurable effect on the underlying perovskite structure, as illustrated in Fig.~\ref{fgr:application}b. Octahedral tilting, quantified by the Pb--X--Pb angle as a deviation from the ideal cubic value of \ang{180}, is a sensitive probe of structural distortion. A previous ab initio study reported average tilt angles of \ang{11} for FAPI and $\approx$~\ang{20} for FA$_{0.9}$Rb$_{0.1}$PbI$_3$.\cite{ghosh_good_2017} Our simulations of Cs$_{0.12}$FA$_{0.88}$PbI$_{0.75}$Br$_{0.25}$ yield an average close to \ang{15}, consistent with these reports. The tilting increases as the surface becomes saturated with AEAPTMS molecules, reaching an average value of $\approx$~\ang{20} for the Pb--X--Pb angle. The increase in octahedral tilting is coupled with an elongation of the Pb--X bonds, linked to the lattice disruption determined by the increased density of AEAPTMS molecules at the surface. This effect is shown in Fig.~\ref{fgr:application}c--e, presenting a top view of the simulated surfaces with a variation in molecular passivator coverage. A decrease in the number of ordered, valid octahedra is observed with an increase in passivation coverage. Here, a PbX$_6$ octahedron is classified as valid if it presents the expected coordination (6-fold in the bulk layers and 5-fold at the surface, with the apical halide missing) and maintains corner-sharing connectivity with neighbouring octahedra. For the single-molecule case (Fig.~\ref{fgr:application}c), all octahedra across the 5 MD repeats retained their expected structural order. At the highest coverage studied (Fig.~\ref{fgr:application}e, 1.28~molecules/nm$^{2}$), our simulations yield an average of 91.5\% $\pm$ 4.2\% valid octahedra, with 7.1\% $\pm$ 3.7\% defective (predominantly undercoordinated) polyhedra and 1.4\% $\pm$ 1.2\% exhibiting edge- or face-sharing connectivity (Fig.~S12). The progressive increase in structural disorder with passivation coverage is in agreement with the experimentally-observed decrease in XRD peak intensities with increasing AEAPTMS deposition time.\cite{huang_passivation_2025} 

The amino-silane molecules can penetrate into the surface, as previously shown by time-of-flight secondary ion mass spectrometry, with AEAPTMS showing comparatively less penetration than related passivators.\cite{bandgap_science} This explains the breaking-down of the octahedral lattice order, as molecules change to an upright orientation with increased passivation coverage, as illustrated in the simulation snapshots in the bottom panels of Fig.~\ref{fgr:application}c--e. As the molecules start reaching below the surface, the corner-sharing connectivity between the octahedra is disturbed (Fig.~\ref{fgr:application}e and Fig.~S13).

Thus, the experimentally observed decline in passivation effectiveness at longer deposition times\cite{huang_passivation_2025} can be linked to the evolution of molecular orientation with passivation coverage. As molecular density increases, steric interactions between methoxy groups of neighbouring molecules disfavour multidentate coordination, instead promoting an upright orientation in which the molecules point downward with the N atoms coordinating to the surface. This is reflected in the elongation of Pb--N bonds and the near-complete disappearance of Pb--O contacts at high coverage (Fig.~S14). The prevalence of upright silane orientations and N-mediated surface coordination is consistent with angle-resolved XPS measurement reports,\cite{jariwala_reducing_2021} and our high-coverage snapshots more realistically mirror experimental deposition conditions than what can be achieved using single-molecule simulations.

\section*{Discussion}

Focusing on the molecular passivation of hybrid perovskite materials, we have explored the role of data in MLIP training for chemically heterogeneous systems. We showed that, where relevant configurations are underrepresented in the pre-training dataset, an approach resembling continual fine-tuning can be useful: integrating large, curated, material-specific datasets as an intermediate step in the training pipeline. Many such datasets are already available,\cite{yuxing_zhou_2023_8208202, ibragimova_2024_10925480, biswas_2025_17363611} and our work has added our `hyP-26' dataset to this growing body of resources. Following fine-tuning on this intermediate dataset (`\textbf{D1}' in Figure~\ref{fgr:overview}a), we then further fine-tune on highly-specialised data (`\textbf{D2}'): an example of where this works well is the \texttt{MACE-MP-0b3} model primarily pre-trained on inorganic structures, \cite{batatia_foundation_2025} which is improved by explicit exposure to the `N1/N2/O' multidentate coordination of the amino-silane molecule. In contrast, the more recently developed \texttt{MACE-MH-1} model \cite{batatia_cross_2025} can already `discover' this type of structure during relaxation, and does not benefit in the same way from multistep fine-tuning. In the longer run, we therefore expect to see convergence in terms of the coverage achieved by pre-training datasets and foundational MLIPs.

From a materials chemistry perspective, our simulations provide atomistic insights into the structural disruption at perovskite surfaces induced by amino-silane molecules -- thus offering a microscopic explanation for the experimentally-observed reduced crystallinity in films and the surface charge blocking effect in devices with excessive deposition time of passivation molecules.\cite{huang_passivation_2025} We had a first look at the dynamic effects at real perovskite surfaces, as we modelled the collective behaviour of an increasing number of passivating molecules: a larger number of steric interactions between methoxysilane groups at higher passivation coverages induce a shift in molecular orientation towards upright configurations. This steric crowding drives the amino-silane molecule to coordinate through its N atoms alone, rather than involving additional O$\rightarrow$Pb binding, and the molecule is able to penetrate into the surface with disordering effects upon the perovskite lattice. Hence, beyond binding efficacy, the design of future molecular passivators ought to take into account steric effects, intermolecular interactions, and deposition-time/molecular coverage effects: for molecular passivators with significant steric bulk, the optimum deposition time should balance maximising favourable passivator--surface interactions with minimising intermolecular interactions. 

A limitation of the present work is our focus on the Pb-terminated (001) surface of the perovskite, without accounting for reactions between the FA$^+$ cations and the amino-silane passivator molecules, or for possible intermolecular coupling of the latter.\cite{jariwala_reducing_2021} A logical next step is therefore to further expand the scope of our molecular-passivation simulations, exploring questions of reactivity and the effect of different defect types at the surface; to this end, the `hyP-26' dataset could be supplemented with relevant configurations from existing datasets containing hybrid perovskite defect structures\cite{Biswas_2026} or reactive organic systems.\cite{Lee_chemrxiv} 

With a long-term view, we anticipate that MLIP-driven simulations can enable the screening of thousands of different molecules that could be used for passivating perovskite films before laboratory tests are carried out. Beyond screening already-known molecules for desirable properties, integrating generative models into the protocol could speed up the design and selection of optimised passivators. Computational materials design has historically focused on bulk crystalline structures, with notable examples for perovskite materials, \cite{Li_predicting_2018, chenebuah_deep_2024, Zhang_PV-design-review_2026} and has begun to extend to the amorphous state.\cite{liu_amorphous_2025, yang_generative_2025} In this context, our study now provides a blueprint for modelling -- and a stepping stone for designing -- heterogeneous multi-component systems in materials chemistry. For solar-cell materials, understanding additives and passivators, and their collective behaviour, is important for improving device operability and durability. Ultimately, a close integration of initial computational screening, property prediction, and experimental verification could therefore enable reliable `digital experiments' guiding the design of photovoltaic devices.

\clearpage

\setstretch{1.2}

\section*{Methods}

\textbf{MLIP fitting.}
All MLIP models reported herein are based on the MACE architecture\cite{Batatia_mace} and pre-trained models,\cite{batatia_foundation_2025} subsequently fine-tuned using the \texttt{graph-pes} software, version 0.2.0.\cite{graph_pes_2024, gardner_2026_18714766} Four different pre-trained models were used, viz.\ \texttt{MACE-MP-0b3}, \texttt{MACE-MPA-0}, \texttt{MACE-OMAT-0}, and \texttt{MACE-MH-1}; these models had been initially trained using the MPTrj,\cite{deng_2023_chgnet} MPTrj+sAlex,\cite{cavignac_ai-driven_2026} OMAT24,\cite{barroso-luque_open_2024} and a combination of the OMAT/OMOL\cite{omol25_2025}/OC20\cite{oc20_2021}/MATPES\cite{kaplan_foundational_2025} datasets, in the stated order (in the notation of Fig.~\ref{fgr:overview}, these are the respective \textbf{D0} datasets). In all cases, we used a `medium' model size, with a maximum degree of equivariance of $L=1$, meaning that the irreducible representations of the messages have alternating parity, \texttt{128x0e + 128x1o}. Details about the hyperparameters are given in Supplementary Note 1. 

\textbf{Reference data.}
The general metal and hybrid perovskite dataset, here denoted `hyP-26', was constructed using a combination of data augmentation techniques, MD snapshots, and  \textit{de novo} exploration\cite{bernstein_novo_2019} using \texttt{autoplex};\cite{liu_automated_2025} details about its full composition and the protocol used in its construction are given in Supplementary Note 2. 
The specialised dataset for modelling molecular passivation, \textbf{D2}, was constructed from a combination of relaxation and NVT MD snapshots. Starting structures were constructed with the passivating molecule [3-(2-aminoethylamino)\-propyl]trimethoxysilane (AEAPTMS) coordinating via its O or N atoms to Cs$_{0.12}$FA$_{0.88}$PbI$_{0.75}$Br$_{0.25}$ surface slabs. Both reference datasets were labelled with density-functional theory (DFT) data, obtained using the projector augmented-wave method \cite{blochl_projector_1994} as implemented in the Vienna Ab Initio Simulation Package (VASP).\cite{kresse_efficiency_1996, kresse_efficient_1996, kresse_ultrasoft_1999} The computations were performed at the SCAN\cite{sun_strongly_2015} level of theory (Supplementary Note 3).

\textbf{MD simulations.}
ML-driven MD simulations (Fig.~\ref{fgr:application}) were carried out using the \texttt{ASE} MACE calculator provided by \texttt{graph-pes}.\cite{gardner_2026_18714766} For each passivation coverage, 5 independent runs were performed at 450~K using the Langevin thermostat implemented in \texttt{ASE}, holding for 500~ps with a timestep of 1~fs. The MD annealing was followed by structural relaxations driven by the \texttt{MACE-MH-1}-derived `FM$\rightarrow$hyP-26$\rightarrow$AEAPTMS' MLIP using the \texttt{LBFGSLineSearch} algorithm, with a force tolerance of 0.075~eV~\AA{}$^{-1}$.  
The simulated structures contained between 1,816 and 3,220 atoms (dependent on molecular coverage), with surface slab dimensions of $\approx$~60$\times$50 \AA{}$^{2}$ and a vacuum size of 50 \AA{} along the $z$-axis (slab models being constructed via \texttt{pymatgen}\cite{Tran2016}). For bond-length calculations, the Pb--X intermolecular distances were averaged over all contacts up to a cut-off of 3.8~\AA{}. The values for the octahedral tilting are reported as $180^{\circ}-\gamma$, where $\gamma$ is the Pb--X--Pb angle along the $x$-axis; for a cubic perovskite phase, the Pb--X--Pb angle along the $x$-axis is taken as representative of the in-plane octahedral tilting, as tilting in the $y$-direction is equivalent by symmetry. 

\section*{Data availability}

Data supporting this work are available at https://github.com/vldgroup/papers-hyP-26.

\section*{Acknowledgements}

We thank Yuxing Zhou for help with \texttt{autoplex} runs, and Litong Wu for useful discussions. L.-B.P. acknowledges funding from the EPSRC Centre for Doctoral Training in Inorganic Chemistry for Future Manufacturing (OxICFM), EP/S023828/1. We are grateful for computational support from the UK national high performance computing service, ARCHER2,\cite{beckett2024archer2} for which access was obtained via the UKCP consortium and funded by EPSRC grant ref EP/X035891/1.

\end{document}


\title{\fontsize{20}{26}\selectfont {\bf Supplementary Information for}\\ `Data-driven atomistic modelling of hybrid halide perovskite passivation'}

\author[1]{Laura-Bianca Pa\c{s}ca}
\author[2]{Henry J. Snaith}
\author[1]{Volker L. Deringer\thanks{E-mail: \url{volker.deringer@chem.ox.ac.uk}}}

\affil[1]{Inorganic Chemistry Laboratory, Department of Chemistry, University of Oxford, \protect\\ Oxford, UK}
\affil[2]{Department of Physics, University of Oxford, Oxford, UK}

\date{}

\maketitle

\clearpage

\setstretch{1.1}

\section*{Supplementary Note 1 (Fine-tuning protocol)}

All fine-tuned MLIPs were obtained using the \texttt{graph-pes} Python package, version 0.2.0.\cite{graph_pes} Four different pre-trained MACE models (\texttt{MACE-MP-0b3}, \texttt{MACE-MPA-0}, \texttt{MACE-OMAT-0}, \texttt{MACE- MH-1}),\cite{Batatia_mace, batatia_cross_2025} all with a `medium' model size, were fine-tuned on the full `hyP-26' dataset. A cut-off radius of 6~\AA{} was used, with weights for energy, force, and virial set to $(\lambda_E, \lambda_F, \lambda_v) = (1, 10, 100)$ in the loss function, which are commonly recommended settings for fine-tuning MACE-based FMs.\cite{Batatia2023arXiv} For the fine-tuning step on the specialised \textbf{D2} dataset, the weights were adjusted to $(\lambda_E, \lambda_F, \lambda_v) = (1, 50, 1)$, as this was found to substantially improve force predictions (a drop in force RMSE on the AEAPTMS test set of around 50\% when compared to an equivalent model fine-tuned using the (1,10,100) hyperparameters). 

We used the Adam optimiser \cite{adam} with the learning rate set to $10^{-4}$; for \texttt{MACE-MH-1}, a value of 10$^{-5}$ was selected when fine-tuning on `hyP-26', as it led to improved performance in the CFT pipeline, assessed by the adsorption energy ($E_{\textnormal{ads}}$) benchmark on the AEAPTMS system. The PyTorch scheduler \texttt{ReduceLROnPlateau} was used to reduce the learning rate by a factor of 0.5 if the monitored loss on the validation set did not improve for 5 consecutive epochs.
Training was performed for a maximum of 100 epochs. Stochastic weight averaging (SWA) was enabled after 80 epochs, during which the energy and force loss weights were increased to $\lambda_E = 1000$ and $\lambda_F = 100$, respectively. Early stopping was applied such that training was terminated if the validation-set accuracy did not improve for 20 consecutive epochs (15 consecutive epochs when fine-tuning on \textbf{D2}), using a force RMSE tolerance of $10^{-3}$~eV/\AA{}. The same training conditions were maintained for the directly-trained model (`direct-hyP-26+AEAPTMS'). 

The test-set energy and force component RMSE values for the `hyP-26'-fine-tuned models (`FM$\rightarrow$\textbf{hyP-26}') are shown in Table \ref{tab:RMSE-FM-hy-PV26}, and those for `FM$\rightarrow$\textbf{AEAPTMS}' and `FM$\rightarrow$\textbf{hyP-26}$\rightarrow$ \textbf{AEAPTMS'} are shown in Table \ref{tab:RMSE-FM-AEAPTMS}.

With the multi-head fine-tuning protocol implemented in \texttt{mace-torch},\cite{batatia_cross_2025} the \texttt{MACE-MP-0b3}-derived `FM$\rightarrow$AEAPTMS' model showed an energy RMSE of 9.9~meV/atom and a force RMSE of 73.8~meV/\AA{}, while the `FM$\rightarrow$hyP-26$\rightarrow$AEAPTMS' model showed an energy RMSE of 10.0~meV/atom and a force RMSE of 71.1~meV/\AA{}; the \texttt{MACE-OMAT-0}-derived `FM$\rightarrow$AEAPTMS' model yielded an energy RMSE of 7.7 meV/atom and a force RMSE of 59.9~meV/\AA{}, while the `FM$\rightarrow$hyP-26$\rightarrow$AEAPTMS' showed energy and force RMSEs of 14.2~meV/atom and 81.6~meV/\AA{}, respectively. We did not have an implementation of the multi-head fine-tuning protocol available when the research for this study was conducted, and as such we did not pursue this direction in the present work.  

\clearpage
\section*{Supplementary Note 2 (Components of the reference datasets)}

\subsection*{Construction of the general `hyP-26' dataset}

The total number of structures and atoms in each class is provided in Table~\ref{tab:Dataset}.

\subsubsection*{Distorted crystalline structures}
For each of the most common metal halide and hybrid halide perovskites (\ce{CsPbI3}, \ce{CsSnI3}, \ce{MAPbI3}, \ce{FAPbI3}), three different crystalline phases (orthorhombic, tetragonal, and cubic) were included in the training dataset, as detailed in Table~\ref{tab:perovskites}.  
Structures were distorted by scaling the unit-cell volume from --7\% to +7\% in increments of 1\% and by varying the cell angles from  --\ang{30} to +\ang{30} in increments of \ang{5} relative to the reference structure. Each volume-scaled structure was additionally rattled using the modified Monte Carlo rattling procedure implemented in \texttt{hiPhive},\cite{eriksson_hiphive} with a displacement standard deviation of 2.5\% and a minimum interatomic distance of 1~\AA{}. The most common crystalline phases of \ce{PbX2} (X = Cl, Br, I) were included following the same protocol, with volume scaling from --5\% to +5\% and angle variations from --\ang{10} to +\ang{10}.

For the hybrid perovskites, additional structures sampling octahedral tilting and molecular cation rotations were included. Octahedral tilts were generated using \texttt{perovskite\_rotater},\cite{islam_normal_2013} which implements the 14 tilting patterns for a cubic perovskite unit cell. For simplicity, only cubic \ce{CsPbI3} was used to generate \ce{PbI6} rotations of up to \ang{40} in increments of \ang{5}. The effect of compositional substitution on octahedral tilting was assumed to be captured by the MD snapshots included elsewhere in the dataset (see `NPT MD snapshots' below), and the influence of the A-site cation on the crystal structure was taken to be primarily steric in nature ($r_\textnormal{Cs}=1.88$ {\AA} $\approx$ $r_\textnormal{MA}=2.17$ {\AA}\cite{oku_crystal_2020}), with limited contribution to the electronic structure.\cite{yin_unusual_2014, giorgi_cation_2014}

Molecular cation rotations were sampled by rigidly rotating the organic cation up to \ang{40} in increments of \ang{5} along each Cartesian axis independently and along all three axes simultaneously, while keeping the inorganic framework fixed. Structures in which any interatomic distance fell below the sum of the crystal radii of the two atoms involved were discarded; the relevant distance thresholds are given in Table~\ref{tab:min-dist}.

\subsubsection*{Site occupancy disordered structures}

Site occupancy disorder modelling, as implemented in \texttt{sod},\cite{Grau-Crespo_2007} was used to create site-disordered structures with mixed-cation or mixed-anion compositions. The generated structures were then each volume-scaled from --2\% to 2\% in increments of 1\%. The mixed-anion structures were additionally MC-rattled with a standard deviation of 5\% using \texttt{hiphive}.\cite{eriksson_hiphive} 

\begin{enumerate}

\item \textit{Mixed-cation structures.} For the (Cs, FA), (Cs, MA), and (MA, FA) mixed-cation compositions, site-disordered supercells were generated from two starting structures: a $2 \times 2 \times 2$ cubic unit cell, with the lattice parameter estimated using Vegard's law,\cite{mcmeekin_crystallization_2017} and the tetragonal unit cell taken from the Inorganic Crystal Structure Database (ICSD),\cite{Fabini_tetragonal_2016} dependent on the A-site cation ratios: for the cubic supercells, cation ratios of 12.5/87.5, 25/75, 50/50, 75/25, and 87.5/12.5 were sampled; for the tetragonal cells, ratios of 25/75, 50/50, and 75/25 were generated.

\item \textit{Mixed-anion structures.} Only the (I, Br) and (Br, Cl) compositions were considered, as the (I, Cl) mixture was found to be experimentally unstable due to the size mismatch between the I$^-$ and Cl$^-$ anions.\cite{unger_chloride_2014} For the CsPbI$_x$Br$_{1-x}$ and CsPbBr$_\textnormal{x}$Cl$_{1-x}$ compositions, the tetragonal cell was used to obtain structures substituted in increments of 16.67, 33.33, 50, 66.67, and 83.33\% of Br substitution. For the hybrid perovskite (I, Br) compositions, both cubic and tetragonal unit cells were used as starting structures. The cubic phase was sampled at Br fractions of 4.2, 8.3, 25, 33.3, and 66.7\%, reflecting the experimental observation that higher Br content stabilises the cubic phase;\cite{jeon_compositional_2015} the tetragonal phase was sampled at 8.3 and 16.7\%. For the (Br, Cl) compositions, only Cl substitutions of 4.2, 8.3, and 16.7\% were considered, consistent with the low Cl doping levels relevant to experimental mixed-anion perovskites.\cite{rev_compositions_2015}

\end{enumerate}

\subsubsection*{Hard-sphere randomized structures}

Randomised structures were generated under hard-sphere (minimum-distance) constraints, each containing 27 A-site cations, 27 B-site cations, and 81 anions, at a density of 4.1~g~cm$^{-3}$ representative of an amorphous phase derived from perovskite materials.\cite{PhysRevB.71.094205} The A-site was populated with MA, FA, and Cs in random proportions,  the B-site with Pb and Sn in random ratios, and the halide sites with I, Br, and Cl, with the Cl content constrained to a maximum of 5\% of the total halide count.

\subsubsection*{GAP-driven random structure search (GAP-RSS) structures}

Random structure search (RSS) structures can be additions to large, generalised datasets which aim to cover a broad range of material phases, including amorphous or polycrystalline materials.\cite{bernstein_novo_2019, JMCA_Bianca} For each of the common iodide crystal phases (\ce{CsPbI3}, \ce{MAPbI3} and \ce{FAPbI3}), we included 1,000 RSS structures obtained across 10 iterations of the GAP-RSS protocol \cite{GAP-RSS_2018, bernstein_novo_2019} implemented in the \texttt{autoplex} framework.\cite{liu_automated_2025} The structures were restricted to between 1 and 8 formula units, with a minimum distance requirement (\#\texttt{MINDIST} (\AA{}): $\textnormal{Cs}-\textnormal{Cs}=4$, $\textnormal{Cs}-\textnormal{Pb}=5$, $\textnormal{Cs}-\textnormal{I}=4$, $\textnormal{Pb}-\textnormal{Pb}=5$, $\textnormal{Pb}-\textnormal{I}=3.04$, $\textnormal{I}-\textnormal{I}=3.5$ for \ce{CsPbI3}, $\textnormal{C}-\textnormal{C}=3.5$, $\textnormal{C}-\textnormal{Pb}=5$, $\textnormal{C}-\textnormal{I}=4$, $\textnormal{Pb}-\textnormal{Pb}=5$, $\textnormal{Pb}-\textnormal{I}=3.04$, $\textnormal{I}-\textnormal{I}=4$, $\textnormal{H}-\textnormal{H}=1.66$, $\textnormal{C}-\textnormal{H}=1.09$, $\textnormal{N}-\textnormal{H}=1.03$, $\textnormal{C}-\textnormal{N}=1.49$, $\textnormal{N}-\textnormal{N}=2$ for \ce{MAPbI3} and \ce{FAPbI3}). 

To describe the atomic environments in the GAP models,\cite{GAP_2010} we used an expansion into a squared-exponential two-body, three-body, and a smooth overlap of atomic positions (SOAP) \cite{bartok_soap} term with maximum number of radial and angular basis functions of $n_{\text{max}}=8$ and $l_{\text{max}}=6$, respectively. The SOAP kernel function was set to a dot product with exponent $\zeta=4$. Details about the hyperparameters of the SOAP descriptor are given in Table~\ref{tab:SOAP-hyper}.

\subsubsection*{NPT MD snapshots}

An MLIP directly-trained on the dataset described above was used to run MD simulations in the NPT ensemble. Three $2 \times 2 \times 2$ cubic supercells were generated using the special quasi-random (SQS) structure approach \cite{SQS_1990} implemented in the \texttt{sqsgenerator}\cite{sqsgen} Python package, with compositions of Cs$_{0.15}$FA$_{0.85}$PbI$_{0.77}$Br$_{0.23}$, Cs$_{0.25}$FA$_{0.75}$PbI$_{0.75}$Br$_{0.21}$Cl$_{0.4}$, and \linebreak Cs$_{0.15}$FA$_{0.6}$MA$_{0.3}$Pb$_{0.5}$Sn$_{0.5}$I$_3$. The unit-cell length was set to 6.24 \AA{}, as expected from the Vegard's-law trend observed experimentally for structures with a similar stoichiometric composition.\cite{mcmeekin_crystallization_2017}
The three different structural models were then used to run NPT simulations starting from 300~K, with an initial annealing step of 10~ps, and then heating up to 600~K over 9~ns. The gradual heating of the three different structures was evenly sampled to select 100 snapshots from each simulation, which were added to the training dataset.

\subsubsection*{Defects}

Vacancies, substitutional and interstitial defects were generated using the workflow implemented in the \texttt{doped} Python package. \cite{Kavanagh2024} Defects were thus generated for \ce{MAPbI3} (vacancies, substitutional and interstitial defects), \ce{CsPbI3}, \ce{CsPbBr3} (only vacancies and substitutional defects) and \ce{CsPbCl3} (only Cl vacancies).

\subsubsection*{Surface structures}

The \texttt{SlabGenerator} function in the \texttt{pymatgen} \cite{pymatgen} \texttt{core} package was used to generate surface slab models representing the (001), (110) and (111) surfaces of \ce{CsPbI3}, \ce{MAPbI3}, \ce{FAPbI3}, 
\ce{Cs$_{0.15}$FA$_{0.85}$PbI$_{0.77}$Br$_{0.23}$},
\ce{Cs$_{0.25}$FA$_{0.75}$PbI$_{0.75}$Br$_{0.21}$Cl$_{0.4}$},  Cs$_{0.15}$FA$_{0.6}$MA$_{0.3}$Pb$_{0.5}$Sn$_{0.5}$I$_3$, \linebreak \ce{PbI2}, \ce{PbBr2}, and \ce{PbCl2}. The minimum slab thickness was 10 {\AA} with a minimum vacuum thickness of 10 {\AA}.

All training and test data were subsequently filtered, removing any structures for which the DFT-calculated forces exceeded 50 eV/\AA{}.

\subsection*{Construction of the domain-specific `\textbf{D2}' dataset}

The \texttt{MACE-MP-0b3} model fine-tuned on `hyP-26' (`FM$\rightarrow$hyP-26') was used to run NVT MD simulations (450~K, 200~ps) using the Langevin thermostat implementation in the ASE Calculator provided by \texttt{graph-pes}.
Starting structures were constructed with the passivating molecule [3-(2-Aminoethylamino)propyl]trimethoxysilane (AEAPTMS) coordinating to Cs$_{0.12}$FA$_{0.88}$PbI$_{0.75}$Br$_{0.25}$ surface slabs via the O or N heteroatoms, respectively. The (001), (110) and (111) surface terminations were considered, using slabs of minimum 10 \AA{} with a vacuum size of 25 \AA{} along the $z$-axis. From each trajectory, 10 equally-spaced snapshots were selected for DFT labelling, with the exception of the larger (111) system, from which only 5 snapshots were used. Thus, 90 structures were labelled in total, with a 9:1 split between the training and test datasets.

After this first iteration, two potentials were trained: the first version used the \texttt{MACE-MP-0b3} foundation model fine-tuned only on the AEAPTMS trajectory snapshots (`FM$\rightarrow$AEAPTMS'), and the second version was a MACE model directly-trained on the full `hyP-26' dataset plus the \textbf{D2} trajectory snapshots (`direct-hyP-26+AEAPTMS'). ML-driven MD simulations were then run with more complex starting structures, which had not been included in the initial training dataset: the AEAPTMS molecule passivating the (001) surface with a Pb vacancy, halide vacancy, and FA--I vacancies, and the (110) surface with an FA vacancy. From each of the trajectories sampled with the two models, 10 evenly-spaced structures were selected for the (001) surface system, and 5 structures each for the larger (110) system. Thus, in the second iteration, an additional 70 structures were labelled, with a 9:1 split between the training and test datasets.

To be able to observe the effect of the passivating molecule on the perovskite's stability to degradation, the model must be able to drive simulations in a higher-temperature regime. To this end, temperature ramps up to 850~K were performed using the second iteration of the `FM$\rightarrow$AEAPTMS' model. The heating rate was set to 2$\times$10$^{12}$~K~s$^{-1}$ for each of the (001), (110), (111), and (200) slab systems with a passivating AEAPTMS molecule at the surface. From each of these trajectories, 15 evenly-spaced snapshots were labelled with DFT and added to the training and test datasets with a 9:1 split.

The fine-tuned \texttt{MACE-MP-0b3} models failed to identify the `N1/N2/O' coordination type as an energy minimum, whereas this posed no difficulty for the fine-tuned  \texttt{MACE-MH-1} models. To enable a consistent comparison of MLIP-relaxed `N1/N2/O' structures across both FMs, the training dataset was augmented for all models with 15 evenly-spaced snapshots drawn from the relaxation trajectories of the molecule--surface system into four coordination types (`N1/N2/O', `N2/O', `N1', and `O vertical', as illustrated in Fig.~1c in the main text) and from the relaxation trajectory of the pristine surface, with relaxations performed using the `direct-hyP-26+AEAPTMS' model. Structural relaxations were carried out with the \texttt{LBFGSLineSearch} algorithm as implemented in ASE, using a maximum force convergence criterion of $\texttt{fmax}=0.075$~eV~\AA{}$^{-1}$. For the multidentate `N1/N2/O' coordination, which proved particularly challenging for the fine-tuned \texttt{MACE-MP-0b3} models, an additional 74 rattled structures were included in the training data.

The test set was expanded with 5 evenly spaced snapshots from each relaxation trajectory obtained using the `FM$\rightarrow$hyP-26$\rightarrow$AEAPTMS' model, supplemented by 25 snapshots drawn from a 500~ps NVT MD trajectory at 350~K.

To test the performance of the MLIP on a related amino-silane passivator not included in the \textbf{D2} dataset, namely trimethoxy[3-(methylamino)propyl]silane (MAPTMS), the 'direct-hyP-26+AEAPTMS' model was used to conduct an NVT MD simulation for 500~ps at 350~K. 50 evenly-spaced snapshots were selected and labelled with the same DFT settings; these were used to test the MLIP force RMSE on an out-of-distribution system, with results shown in Fig.~\ref{fgr:RMSE-MAPTMS}.

\clearpage
\section*{Supplementary Note 3 (DFT computations)}

All DFT calculations were performed at the meta-GGA SCAN\cite{sun_strongly_2015} level of theory, using the Vienna Ab Initio Simulation Package (VASP).\cite{kresse_ab_1993, kresse_efficiency_1996, kresse_efficient_1996}  The projector augmented-wave (PAW) pseudopotentials\cite{blochl_projector_1994, kresse_ultrasoft_1999} used corresponded to valence-electron configurations of 4s$^2$4p$^5$ for Br, 2s$^2$2p$^2$ for C, 3s$^2$3p$^5$ for Cl, 5s$^2$5p$^6$6s$^1$ for Cs (\texttt{Cs\_sv}), 1s$^1$ for H, 5s$^2$5p$^5$ for I, 2s$^2$2p$^3$ for N, 2s$^2$2p$^4$ for O, 5s$^2$5p$^6$5d$^{10}$6s$^2$6p$^2$ for Pb (\texttt{Pb\_sv\_GW}), 3s$^2$3p$^2$ for Si, 4d$^{10}$5s$^2$5p$^2$ for Sn (\texttt{Sn\_d}).
For the single-point calculations, an energy cut-off of 600~eV was used with a stopping criterion of $1\times10^{-6}$~eV and a Monkhorst–Pack grid centred on $\Gamma$ with a $k$-spacing of 0.2~\AA{}. Gaussian smearing was applied with a width of 0.05~eV. 

Our calculations were performed without the \texttt{LASPH} setting, which includes non-spherical contributions to the corrections for the gradient of the density inside the PAW spheres.\cite{VASP_manual} As this setting was found to affect the single-point energies of related BiMO$_3$ perovskite systems by up to $\approx 6$~meV/atom,\cite{LASPH_2022} we have tested the effect of setting \texttt{LASPH} to \texttt{True} and retraining MLIPs on the thus-relabelled \textbf{D2} dataset; the results are shown in Table~\ref{tab:mlip_adsorption_LASPH}. The largest difference is observed for the isolated AEAPTMS molecule: $\approx$6~meV/atom between the $\Delta E$ (with respect to DFT) of the MLIP trained on the dataset with \texttt{LASPH} turned off and that of the MLIP trained on the dataset with \texttt{LASPH} turned on. 
We checked the performance of models trained on the \textbf{D2} dataset labelled with \texttt{LASPH}=True and found that it does not change the trend observed in Fig.~3; hence we deem the original DFT settings used for our reported results sufficient to support the message in the main text.

For the geometry relaxations, the stopping criterion was decreased to 1 $\times$ 10$^{-5}$~eV for the energy and a setting of --0.075~eV/\AA{} for \texttt{EDIFFG}, stopping the relaxation when the norms of all the forces are all lower than $\mid\texttt{EDIFFG}\mid$. The relaxations were started from the lowest-energy MLIP-relaxed structure for each configuration type.

\clearpage

\section*{Supplementary Tables}

\begin{table*}[h!]
\centering
\caption{Composition of the training dataset.}
{\small
\begin{tabular}{llccc}
\toprule
\multicolumn{2}{l}{} & \multicolumn{2}{c}{Dataset size} \\
\cmidrule(rl){3-4}
 {Training data} &  & {Cells} & {Atoms}\\
\midrule
Free atoms & Cs/C/H/N/Pb/Sn/Cl/Br/I & 9 & 9 \\
\hline
& & & \\
\multirow{14}{*}{Crystalline structures} & \ce{CsPbI3}  & 383 & 7,535 \\
 & \ce{FAPbI3} & 1,101 & 64,393  \\
 & \ce{MAPbI3} & 540 &  65,808  \\
 & \ce{CsSnI3} & 467 & 21,395   \\
 & \ce{Cs$_\textnormal{x}$FA$_{1-\textnormal{x}}$PbI3} & 85  & 5,187  \\
 & \ce{Cs$_\textnormal{x}$MA$_{1-\textnormal{x}}$PbI3} & 85 & 5,005 \\
 & \ce{MA$_{\textnormal{x}}$FA$_{1-\textnormal{x}}$PbI$_3$} & 85 & 7,202 \\
 & \ce{CsPb$_\textnormal{x}$Sn$_{1-\textnormal{x}}$I3} & 33 &  660 \\
& \ce{CsPbBr$_{\textnormal{x}}$Cl$_{1-\textnormal{x}}$}  & 1,721 & 34,420  \\
& \ce{CsPbI$_\textnormal{x}$Br$_{1-\textnormal{x}}$} & 1,721 & 34,420  \\
& \ce{FAPbBr$_\textnormal{x}$Cl$_{1-\textnormal{x}}$} & 108 & 6,912 \\
& \ce{FAPbI$_{\textnormal{x}}$Br$_{1-\textnormal{x}}$} & 189 & 10,392 \\
& \ce{MAPbI$_{\textnormal{x}}$Br$_{1-\textnormal{x}}$} & 153 & 6,984 \\
& \ce{MAPbBr$_{\textnormal{x}}$Cl$_{1-\textnormal{x}}$} & 108 & 6,912 \\
& \ce{PbCl2} & 43 & 3,216 \\
& \ce{PbBr2} & 21 & 1,008 \\
& \ce{PbI2} & 64 & 4,656 \\
&  &  &  \\
\hline 
\multirow{3}{*}{NPT MD snapshots} & \ce{Cs$_{0.15}$FA$_{0.85}$PbI$_{0.77}$Br$_{0.23}$} & 270 & 24,119 \\
& \ce{Cs$_{0.25}$FA$_{0.75}$PbI$_{0.75}$Br$_{0.21}$Cl$_{0.4}$} & 270 & 22,140 \\ & \ce{Cs$_{0.15}$FA$_{0.6}$MA$_{0.3}$Pb$_{0.5}$Sn$_{0.5}$I$_{3}$} & 270  & 49,950 \\
\hline
Random structures (hard-sphere) & Random ABX$_3$ composition & 33 & 6,940  \\
\hline
\multirow{4}{*}{Defect structures} & \ce{CsPbI3} & 77 & 10,356 \\
& \ce{MAPbI3} & 166 & 53,767  \\
& \ce{CsPbBr3} & 73 &  9,836\\ 
& \ce {CsPbCl3} & 13 & 1,742 \\
\hline
\multirow{9}{*}{Surface slabs} & 
\ce{CsPbI3} & 13 & 175 \\ 
& \ce{MAPbI3} & 9 & 276 \\
& \ce{FAPbI3} & 4 & 132 \\
& \ce{Cs$_{0.15}$FA$_{0.85}$PbI$_{0.77}$Br$_{0.23}$} & 14 & 2,143 \\
& \ce{Cs$_{0.25}$FA$_{0.75}$PbI$_{0.75}$Br$_{0.21}$Cl$_{0.4}$} & 11 & 1,558\\
&\ce{Cs$_{0.15}$FA$_{0.6}$MA$_{0.3}$Pb$_{0.5}$Sn$_{0.5}$I$_{3}$} & 28 & 7,770 \\
& \ce{PbCl2} & 22 & 1,044 \\
& \ce{PbBr2} & 9 & 288\\
& \ce{PbI2} & 14 & 546 \\
\hline
\multirow{3}{*}{Random structures (GAP-RSS)} & \ce{CsPbI3} & 1,000 & 8,500 \\
& \ce{MAPbI3} & 1,000 & 23,136 \\
& \ce{FAPbI3} & 1,000 & 24,036 \\
\hline
\textbf{Total} & & \textbf{11,020} &  \textbf{522,723} \\
\bottomrule
\label{tab:Dataset}
\end{tabular}
}
\end{table*}

\begin{table}[h!]
    \centering
    \caption{Phases and space groups of the main perovskite structures included in the training dataset.}
    \renewcommand{\arraystretch}{1.2}
    \begin{tabular}{l c c}
        \hline
        Compound & Phase & Space Group \\
        \hline
        \ce{CsPbI3}  & Cubic ($\alpha$)        & \textit{$Pm\bar{3}m$} (221) \\
                     & Tetragonal ($\beta$)   & \textit{P4/mbm} (127) \\
                     & Orthorhombic ($\gamma$) & \textit{Pnma} (62) \\

        \ce{CsSnI3}  & Cubic ($\alpha$)        & \textit{$Pm\bar{3}m$} (221) \\
                     & Orthorhombic ($\gamma$) & \textit{Pnma} (62) \\
                     & Tetragonal ($\beta$)   & \textit{P4/mbm} (127) \\
        \ce{MAPbI3}  & Cubic\cite{Brivio_hybrid_perovskites_2013}            & \textit{$Pm\bar{3}m$} (221) \\
                     & Tetragonal       & \textit{I4/mcm} (140) \\
                     & Orthorhombic     & \textit{Pnma} (62) \\
        \ce{FAPbI3}  & Cubic ($\alpha$)\cite{Weller_cubic_FAPI_2015}  & \textit{$Pm\bar{3}m$} (221) \\
                     & Tetragonal ($\beta$)  & \textit{P4$_2$/mnm} (136) \\
                    & Hexagonal ($\delta$)    & \textit{P63mc} (186) \\
        \ce{PbCl2} & Monoclinic  & \textit{$P12_1/c1$} \\
                   & Orthorhombic & \textit{$Pnma$} (62) \\
        \ce{PbBr2}  & Orthorhombic & \textit{$Pnma$} (62) \\
        \ce{PbI2}  & Hexagonal & \textit{$P6_3mc$} (186) \\
                   & Trigonal & \textit{$R\bar{3}m$} (160) \\
                   & Trigonal & \textit{$P3m1$} (164) \\
                   
        \hline
    \end{tabular}
    \label{tab:perovskites}
\end{table}

\begin{table}[h!]
    \centering
     \caption{Minimum distance constraints imposed when distorting the crystalline unit cells for data augmentation.}
    \renewcommand{\arraystretch}{1.2}
    \label{tab:min-dist}
    \begin{tabular}{l c}
        \toprule
        Bond & Minimum distance\\
        \midrule
         C--C   & 1.0 {\AA} \\
         C--H  & 0.5 {\AA}  \\
         C--I   & 1.9 {\AA} \\
         C--N & 1.0 {\AA}  \\
         C--Pb  & 2 {\AA}  \\
         H--H & 0.5 {\AA}  \\
         H--I & 1.4 {\AA}  \\
         H--N & 0.5 {\AA}  \\
         H--Pb & 1.4 {\AA} \\
         I--I & 2.2 {\AA} \\
         I--N & 1.9 {\AA} \\
         I--Pb & 2.4 {\AA} \\
         N--N & 1.2 {\AA} \\
         N--Pb & 1.8 {\AA} \\
         Pb--Pb & 2.4 {\AA} \\
        \bottomrule
    \end{tabular}
\end{table}

\begin{table}[h!]
\centering
\caption{SOAP hyperparameters chosen for the GAP-RSS protocol.}
\begin{tabular}{lccc}
\toprule
Hyperparameter & 2-body & 3-body & SOAP\\
\midrule
\texttt{$\delta$} & 2 & 2 & 1  \\
\texttt{cutoff} & 5.0 & 3.25 & 5.0  \\
\texttt{n\_sparse} & 15 & 100 & 1000  \\
\bottomrule
\end{tabular}
\label{tab:SOAP-hyper}
\end{table}

\begin{table}[h!]
    \centering
    \caption{Energy and force RMSEs on the `hyP-26' test set for the `FM$\rightarrow$hyP-26' fine-tuned models, starting from the four different pre-trained MACE foundation models. Values are rounded to two decimal places.}
    \renewcommand{\arraystretch}{1.2}
    \begin{tabular}{l c c}
        \toprule
        `FM$\rightarrow$hyP-26' model & Energy RMSE (eV/atom) & Force RMSE (eV/\AA{}) \\
        \midrule
        \texttt{MACE-MP-0b3}  & 0.01  & 0.11  \\
        \texttt{MACE-MPA-0}  & 0.01   & 0.10  \\
        \texttt{MACE-OMAT-0}  &  0.01    &  0.09 \\
        \texttt{MACE-MH-1}  & 0.01    & 0.07  \\
        \bottomrule
    \end{tabular}
    \label{tab:RMSE-FM-hy-PV26}
\end{table}

\begin{table}[h!]
    \centering
    \caption{Energy and force RMSEs on the AEAPTMS (\textbf{D2}) test set for the `FM$\rightarrow$AEAPTMS' and `FM$\rightarrow$hyP-26$\rightarrow$AEAPTMS' fine-tuned models, starting from four different pre-trained MACE foundation models. Values are rounded to two decimal places.}
    \renewcommand{\arraystretch}{1.2}
    \begin{tabular}{l c c}
        \hline
        `FM$\rightarrow$AEAPTMS' model & Energy RMSE (eV/atom) & Force RMSE (eV/\AA{}) \\
        \hline
        \texttt{MACE-MP-0b3}  & 0.01    & 0.09 \\
        \texttt{MACE-MPA-0}  & 0.01  & 0.08 \\
        \texttt{MACE-OMAT-0}  &  0.01  &  0.08 \\
        \texttt{MACE-MH-1}  &  0.01  & 0.06  \\
        \hline
        `FM$\rightarrow$hyP-26$\rightarrow$AEAPTMS' model & Energy RMSE (eV/atom) & Force RMSE (eV/\AA{}) \\
        \hline
        \texttt{MACE-MP-0b3}  & 0.01   & 0.09 \\
        \texttt{MACE-MPA-0}  & 0.01  & 0.08 \\
        \texttt{MACE-OMAT-0}  &  0.01    & 0.07  \\
        \texttt{MACE-MH-1}  & 0.01   & 0.05 \\
        \hline
    \end{tabular}
    \label{tab:RMSE-FM-AEAPTMS}
\end{table}

\begin{table}[!h]
\caption{Single-point energy differences between MLIP and DFT values computed with different
MLIP fitting approaches using the \texttt{MACE-MP-0b3} and \texttt{MACE-MH-1} foundation models.
Values are reported as $\Delta E = \langle E_{\mathrm{MLIP}} \rangle - E_{\mathrm{DFT}} \pm \sigma$,
where $\sigma$ is the standard deviation over three independent MLIP runs.
All evaluated structures were relaxed using DFT.}
\label{tab:mlip_adsorption}
\centering
\begin{tabular}{lccc}
\toprule
 & \multicolumn{3}{c}{$\Delta E$ (meV/atom)} \\
 \cmidrule{2-4}
Structure & Zero-shot FM & FM$\rightarrow$AEAPTMS & FM$\rightarrow$hyP-26$\rightarrow$AEAPTMS \\
\midrule
\multicolumn{4}{l}{\texttt{MACE-MP-0b3}} \\
\midrule
AEAPTMS & 17.8 $\pm$ 0.0 & 83.4 $\pm$ 2.5 & 11.6 $\pm$ 1.8 \\
Surface & 43.3 $\pm$ 0.0 & -12.0 $\pm$ 0.5 & 0.7 $\pm$ 1.0 \\
N1 & 40.7 $\pm$ 0.0 & 3.1 $\pm$ 0.7 & 1.6 $\pm$ 1.0 \\
N1/N2/O & 44.5 $\pm$ 0.0 & 3.9 $\pm$ 0.7 & 2.2 $\pm$ 1.0 \\
N2/O & 43.8 $\pm$ 0.0 & 4.2 $\pm$ 0.7 & 2.4 $\pm$ 1.0 \\
O vertical & 41.1 $\pm$ 0.0 & 2.9 $\pm$ 0.6 & 1.5 $\pm$ 1.1 \\
\midrule
\multicolumn{4}{l}{\texttt{MACE-MH-1}} \\
\midrule
AEAPTMS & 2.7 $\pm$ 0.0 & 4.8 $\pm$ 2.4 & 8.4 $\pm$ 2.0 \\
Surface & 12.8 $\pm$ 0.0 & 1.0 $\pm$ 0.5 & 0.6 $\pm$ 0.6 \\
N1 & 11.6 $\pm$ 0.0 & 0.4 $\pm$ 0.3 & 0.6 $\pm$ 0.2 \\
N1/N2/O & 13.7 $\pm$ 0.0 & 0.5 $\pm$ 0.3 & 0.7 $\pm$ 0.2 \\
N2/O & 12.5 $\pm$ 0.0 & 0.5 $\pm$ 0.3 & 0.8 $\pm$ 0.2 \\
O vertical & 10.8 $\pm$ 0.0 & 0.7 $\pm$ 0.3 & 1.0 $\pm$ 0.2 \\
\bottomrule
\end{tabular}
\end{table}

\begin{table}[!h]
\caption{Single-point energy differences between MLIP and DFT values computed with different MLIP fitting approaches using the \texttt{MACE-MP-0b3} and \texttt{MACE-MH-1} foundation models, comparing the effects of turning the \texttt{LASPH} setting in VASP off vs.\ on. Values are reported as $\Delta E = \langle E_{\mathrm{MLIP}} \rangle - E_{\mathrm{DFT}} \pm \sigma$ in meV/atom, where $\sigma$ is the standard deviation over three independent MLIP runs. All evaluated structures were relaxed using DFT.}
\label{tab:mlip_adsorption_LASPH}
\centering
\begin{tabular}{lcccc}
\toprule
 & \multicolumn{2}{c}{FM$\rightarrow$AEAPTMS} & \multicolumn{2}{c}{FM$\rightarrow$hyP-26$\rightarrow$AEAPTMS} \\
\cmidrule(lr){2-3} \cmidrule(lr){4-5}
Structure & w/o \texttt{LASPH} & w/ \texttt{LASPH} & w/o \texttt{LASPH} & w/ \texttt{LASPH} \\
\midrule
\multicolumn{5}{l}{\texttt{MACE-MP-0b3}} \\
\midrule
AEAPTMS    &   83.4 $\pm$ 2.5 &   77.5 $\pm$ 2.7 &   11.6 $\pm$ 1.8 &   12.8 $\pm$ 1.5 \\
Surface    &  -12.0 $\pm$ 0.5 &  -13.5 $\pm$ 0.6 &    0.7 $\pm$ 1.0 &    0.4 $\pm$ 0.6 \\
N1         &    3.1 $\pm$ 0.7 &    0.7 $\pm$ 0.8 &    1.6 $\pm$ 1.0 &    1.3 $\pm$ 0.7 \\
N1/N2/O    &    3.9 $\pm$ 0.7 &    1.7 $\pm$ 0.8 &    2.2 $\pm$ 1.0 &    2.1 $\pm$ 0.7 \\
N2/O       &    4.2 $\pm$ 0.7 &    1.9 $\pm$ 0.8 &    2.4 $\pm$ 1.0 &    2.2 $\pm$ 0.7 \\
O vertical &    2.9 $\pm$ 0.6 &    0.6 $\pm$ 0.8 &    1.5 $\pm$ 1.1 &    1.2 $\pm$ 0.7 \\
\midrule
\multicolumn{5}{l}{\texttt{MACE-MH-1}} \\
\midrule
AEAPTMS    &    4.8 $\pm$ 2.4 &    6.3 $\pm$ 0.9 &    8.4 $\pm$ 2.0 &    8.6 $\pm$ 0.8 \\
Surface    &    1.0 $\pm$ 0.5 &    0.3 $\pm$ 0.3 &    0.6 $\pm$ 0.6 &   -0.4 $\pm$ 0.4 \\
N1         &    0.4 $\pm$ 0.3 &   -0.0 $\pm$ 0.4 &    0.6 $\pm$ 0.2 &   -0.1 $\pm$ 0.4 \\
N1/N2/O    &    0.5 $\pm$ 0.3 &    0.3 $\pm$ 0.4 &    0.7 $\pm$ 0.2 &    0.3 $\pm$ 0.4 \\
N2/O       &    0.5 $\pm$ 0.3 &    0.2 $\pm$ 0.4 &    0.8 $\pm$ 0.2 &    0.2 $\pm$ 0.3 \\
O vertical &    0.7 $\pm$ 0.3 &    0.3 $\pm$ 0.3 &    1.0 $\pm$ 0.2 &    0.2 $\pm$ 0.4 \\
\bottomrule
\end{tabular}
\end{table}

\begin{table}[!h]
\caption{Single-point energy differences between MLIP and DFT values computed with different
MLIP fitting approaches using the \texttt{MACE-MPA-0} and \texttt{MACE-OMAT-0} foundation models.
Values are reported as $\Delta E = \langle E_{\mathrm{MLIP}} \rangle - E_{\mathrm{DFT}} \pm \sigma$,
where $\sigma$ is the standard deviation over three independent MLIP runs.
All evaluated structures were relaxed using DFT.}
\label{tab:mlip_adsorption_omat_mpa}
\centering
\begin{tabular}{lccc}
\toprule
 & \multicolumn{3}{c}{$\Delta E$ (meV/atom)} \\
 \cmidrule{2-4}
Structure & Zero-shot FM & FM$\rightarrow$AEAPTMS & FM$\rightarrow$hyP-26$\rightarrow$AEAPTMS \\
\midrule
\multicolumn{4}{l}{\texttt{MACE-MPA-0}} \\
\midrule
AEAPTMS & 29.7 $\pm$ 0.0 & 52.5 $\pm$ 1.4 & 11.0 $\pm$ 0.4 \\
Surface & 0.8 $\pm$ 0.0 & -8.1 $\pm$ 0.9 & 1.0 $\pm$ 0.5 \\
N1 & 5.6 $\pm$ 0.0 & 2.1 $\pm$ 1.0 & 2.7 $\pm$ 0.4 \\
N1/N2/O & 7.7 $\pm$ 0.0 & 3.1 $\pm$ 1.0 & 3.3 $\pm$ 0.5 \\
N2/O & 7.7 $\pm$ 0.0 & 3.2 $\pm$ 1.0 & 3.6 $\pm$ 0.4 \\
O vertical & 5.6 $\pm$ 0.0 & 1.9 $\pm$ 0.9 & 2.6 $\pm$ 0.4 \\
\midrule
\multicolumn{4}{l}{\texttt{MACE-OMAT-0}} \\
\midrule
AEAPTMS & 66.4 $\pm$ 0.0 & 57.7 $\pm$ 0.3 & 9.8 $\pm$ 1.9 \\
Surface & 69.1 $\pm$ 0.0 & -7.8 $\pm$ 1.9 & 1.1 $\pm$ 0.9 \\
N1 & 65.3 $\pm$ 0.0 & 1.4 $\pm$ 1.7 & 1.1 $\pm$ 1.0 \\
N1/N2/O & 66.1 $\pm$ 0.0 & 1.3 $\pm$ 1.7 & 1.1 $\pm$ 1.0 \\
N2/O & 65.1 $\pm$ 0.0 & 1.5 $\pm$ 1.7 & 1.3 $\pm$ 1.0 \\
O vertical & 66.8 $\pm$ 0.0 & 2.6 $\pm$ 1.6 & 2.6 $\pm$ 0.9 \\
\bottomrule
\end{tabular}
\end{table}

\begin{table}[h!]
\caption{Adsorption energies computed with `direct-hyP-26+AEAPTMS'. Values are reported as $\Delta E = \langle E_{\mathrm{MLIP}} \rangle - E_{\mathrm{DFT}} \pm \sigma$, where $\sigma$ is the standard deviation over three independent MLIP runs. The evaluated structures were relaxed using DFT.}
\label{tab:mlip_adsorption_direct}
\center
\begin{tabular}{lc}
\toprule
Structure & direct-hyP-26+AEAPTMS $\Delta E$ (meV/atom)  \\
\midrule
AEAPTMS & 21 $\pm$ 2.0 \\
Surface & 0.90 $\pm$ 0.12 \\
N1/N2/O & 4.5 $\pm$ 0.72  \\
N2/O & 4.6 $\pm$ 0.51 \\
N1 & 3.9 $\pm$ 0.45  \\
O vertical & 3.2 $\pm$ 0.29 \\
\bottomrule
\end{tabular}
\end{table}

\begin{table}[h!]
\caption{
Energy differences, $\Delta E$, between the DFT single-point energy predictions for structures relaxed using DFT and the same structures relaxed with MLIPs for the AEAPTMS passivation test cases. Values are reported as $\Delta E \pm \sigma$, where $\sigma$ is the standard deviation obtained from three independently trained MLIPs using different random seeds. The values are provided in eV for the whole structure, rather than per-atom energies, to allow comparison with $E_\textnormal{ads}$ predictions for the purpose of identifying error cancellation.}
\label{tab:dft_mlip_deltaE}
\renewcommand{\arraystretch}{1.3}
\centering
\resizebox{0.75\textwidth}{!}{
\begin{tabular}{llcc}
\toprule
FM & Structure
& FM$\rightarrow$\textbf{D2} $\Delta E$ (eV)
& FM$\rightarrow$\textbf{D1}$\rightarrow$\textbf{D2} $\Delta E$ (eV) \\
\midrule
\multirow{6}{*}{\texttt{MACE-MP-0b3}}
& N1/N2/O         & 1.27 $\pm$ 0.02 & 0.92 $\pm$ 0.19 \\
& N2/O            &  1.02 $\pm$ 0.16 & 0.87 $\pm$ 0.14 \\
& N1              &  0.24 $\pm$ 0.02 & 0.41 $\pm$ 0.05 \\
& O (vertical)    &  0.38 $\pm$ 0.07 & 0.46 $\pm$ 0.35 \\
& Surface         &  1.72 $\pm$ 0.03 & --0.32 $\pm$ 0.15 \\
& AEAPTMS   & 0.01 $\pm$ 0.00 & 0.01 $\pm$ 0.00 \\
\midrule
\multirow{6}{*}{\texttt{MACE-MH-1}}
& N1/N2/O         &  0.07$\pm$ 0.01 & 0.24 $\pm$ 0.07 \\
& N2/O            &  0.19 $\pm$ 0.10 & 0.11 $\pm$ 0.04 \\
& N1              &  0.19 $\pm$ 0.2 & 0.14 $\pm$ 0.01 \\
& O (vertical)    &  0.07 $\pm$ 0.23 & 0.15 $\pm$ 0.07 \\
& Surface         & --0.09 $\pm$ 0.37 & 0.05 $\pm$ 0.40 \\
& AEAPTMS  &  0.01 $\pm$ 0.005 & 0.01 $\pm$ 0.01 \\
\bottomrule
\end{tabular}
}
\end{table}

\begin{table}[h!]
\caption{
Energy differences $\Delta E$ between the DFT single-point energy predictions for structures relaxed using DFT and the same structures relaxed with MLIPs for the \textbf{D2} (AEAPTMS passivation) test dataset.
Values are reported as $\Delta E \pm \sigma$, where $\sigma$ is the standard deviation obtained from three independently trained MLIPs using different random seeds. The values are provided in eV for the whole structure, rather than per-atom energies, to allow comparison with $E_\textnormal{ads}$ predictions for the purpose of identifying error cancellation.
}
\label{tab:dft_mlip_deltaE_omat_mpa}
\renewcommand{\arraystretch}{1.3}
\centering
\resizebox{0.75\textwidth}{!}{
\begin{tabular}{llcc}
\toprule
FM & Structure
& FM$\rightarrow$\textbf{D2} $\Delta E$ (eV)
& FM$\rightarrow$\textbf{D1}$\rightarrow$\textbf{D2} $\Delta E$ (eV) \\
\midrule
\multirow{6}{*}{\texttt{MACE-MPA-0}}
& N1/N2/O         &  1.10 $\pm$ 0.18  & 0.70 $\pm$ 0.08 \\
& N2/O            &  0.67 $\pm$ 0.30 & 1.08 $\pm$ 0.20 \\
& N1              & 0.41 $\pm$ 0.17 & 0.35 $\pm$ 0.17  \\
& O (vertical)    &   0.34 $\pm$ 0.06 & 0.32 $\pm$ 0.03 \\
& Surface         &  --0.11 $\pm$ 0.44 & --0.51 $\pm$ 0.15 \\
& AEAPTMS   &  0.01 $\pm$ 0.003 & 0.01 $\pm$ 0.003  \\
\midrule
\multirow{6}{*}{\texttt{MACE-OMAT-0}}
& N1/N2/O         &  0.91 $\pm$ 0.01 & 0.83 $\pm$ 0.08   \\
& N2/O            &  0.69 $\pm$ 0.26 & 0.18 $\pm$ 0.15 \\
& N1              &  0.48 $\pm$ 0.2 & 0.35 $\pm$ 0.05  \\
& O (vertical)    &  0.1 $\pm$ 0.41 & 0.4 $\pm$ 0.1 \\
& Surface         &  --0.18 $\pm$ 0.31 & --0.43 $\pm$ 0.01 \\
& AEAPTMS  &  0.01 $\pm$ 0.005 & 0.01 $\pm$ 0.005  \\
\bottomrule
\end{tabular}
}
\end{table}

\begin{table}[h!]
    \centering
    \caption{Comparison of passivated surface systems relaxed using the different potential-fitting approaches, starting from different foundation models. The SOAP similarity score reflects how similar the structure is to the corresponding ground-truth DFT-relaxed configuration (0: completely dissimilar; 1: identical).}
    \renewcommand{\arraystretch}{1.3}
    \resizebox{\textwidth}{!}{
    \begin{tabular}{ll ccc ccc ccc ccc c}
    \toprule
    & & \multicolumn{3}{c}{\texttt{MACE-MP-0b3}} & \multicolumn{3}{c}{\texttt{MACE-MH-1}} & \multicolumn{3}{c}{\texttt{MACE-MPA-0}} & \multicolumn{3}{c}{\texttt{MACE-OMAT-0}} & Direct \\
    \cmidrule(lr){3-5} \cmidrule(lr){6-8} \cmidrule(lr){9-11} \cmidrule(lr){12-14} \cmidrule(lr){15-15}
    \textbf{Config.} & & $\rightarrow$D2 & $\rightarrow$D1$\rightarrow$D2 & zero-shot
                     & $\rightarrow$D2 & $\rightarrow$D1$\rightarrow$D2 & zero-shot
                     & $\rightarrow$D2 & $\rightarrow$D1$\rightarrow$D2 & zero-shot
                     & $\rightarrow$D2 & $\rightarrow$D1$\rightarrow$D2 & zero-shot
                     & D1+D2 \\
    \midrule
    N1/N2/O   & & 0.807 & 0.840 & --    & 0.999 & 0.955 & --    & 0.831 & 0.836 & --    & 0.838 & 0.852 & --    & 0.828 \\
    N2/O      & & 0.836 & 0.861 & 0.724 & 0.998 & 0.979 & 0.833 & 0.819 & 0.820 & 0.740 & 0.844 & 0.819 & 0.794 & 0.850 \\
    N1        & & 0.912 & 0.905 & 0.768 & 0.999 & 0.982 & 0.875 & 0.870 & 0.884 & 0.781 & 0.878 & 0.884 & 0.780 & 0.884 \\
    O vertical& & 0.850 & 0.958 & 0.757 & 0.904 & 0.899 & 0.856 & 0.852 & 0.851 & 0.810 & 0.865 & 0.859 & 0.757 & 0.999 \\
    \midrule
    \textbf{Avg.} & & \textbf{0.851} & \textbf{0.891} & \textbf{0.746}
                  & \textbf{0.975} & \textbf{0.954} & \textbf{0.855}
                  & \textbf{0.843} & \textbf{0.848} & \textbf{0.777}
                  & \textbf{0.856} & \textbf{0.853} & \textbf{0.777}
                  & \textbf{0.858} \\
    \bottomrule
    \end{tabular}
    }
    \label{tab:SOAP-similarity}
\end{table}

\clearpage

\section*{Supplementary Figures}

\begin{figure}[H]
\centering
  \includegraphics[width=0.5\textwidth]{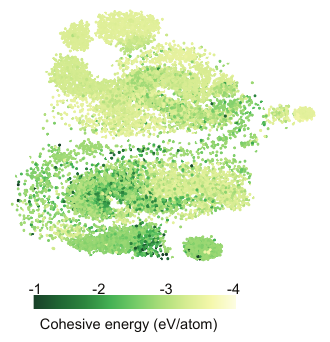}
  \caption{The energetic spread of the `hyP-26' dataset, visualised similar to Fig.~1b in the main text, but now colour-coded according to the cohesive energy of the respective structures. The two-dimensional representation of the configurational space is constructed with the UMAP algorithm \cite{mcinnes_umap_2018} based on the SOAP similarity metric,\cite{bartok_soap} showing the clustering of structurally-similar systems included in the dataset. }
  \label{fgr:SI-UMAP}
\end{figure}

\begin{figure}[H]
\centering
  \includegraphics[width=\textwidth]{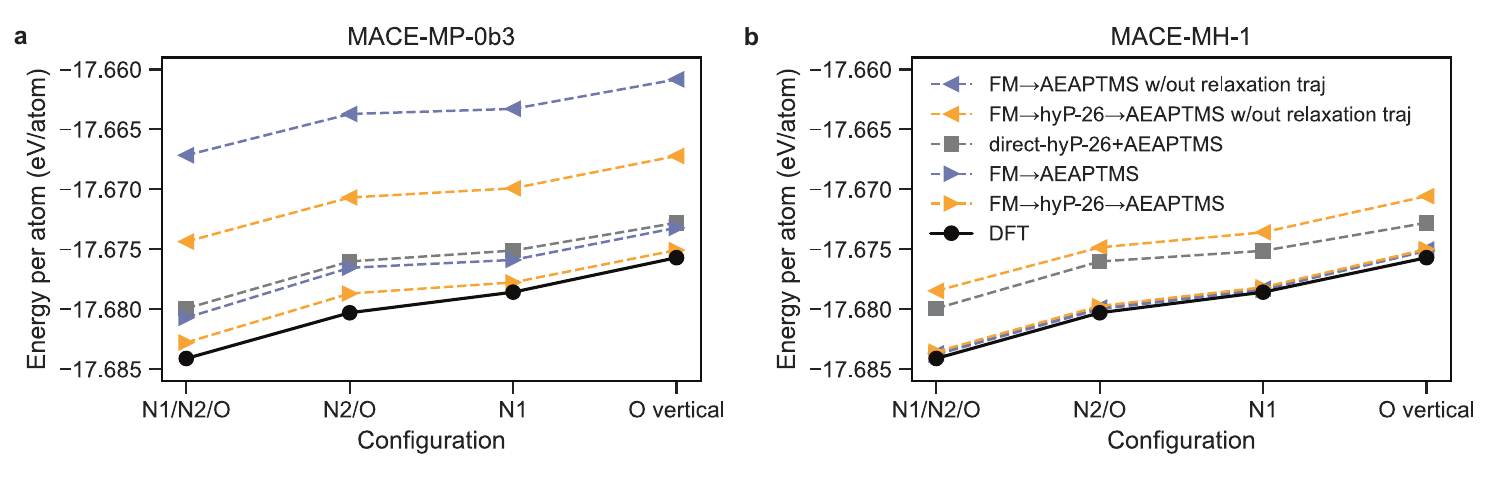}
\caption{MLIP single-point energy predictions for the different configuration types, showing the effect of including snapshots from relaxation trajectories into the four desired configuration types (`N1/N2/O', `N2/O', `N1' and `O vertical', shown in Fig.~1 in the main text). Panel \textbf{a} shows results for \texttt{MACE-MP-0b3}-derived models, while panel \textbf{b} shows results for \texttt{MACE-MH-1}-derived models.   The directly-trained model (`direct-hyP-26+AEAPTMS') predictions and the DFT results are provided for reference.
}
  \label{fgr:SI-relax-inclusion}
\end{figure} 

\begin{figure}[H]
\centering
  \includegraphics[width=\textwidth]{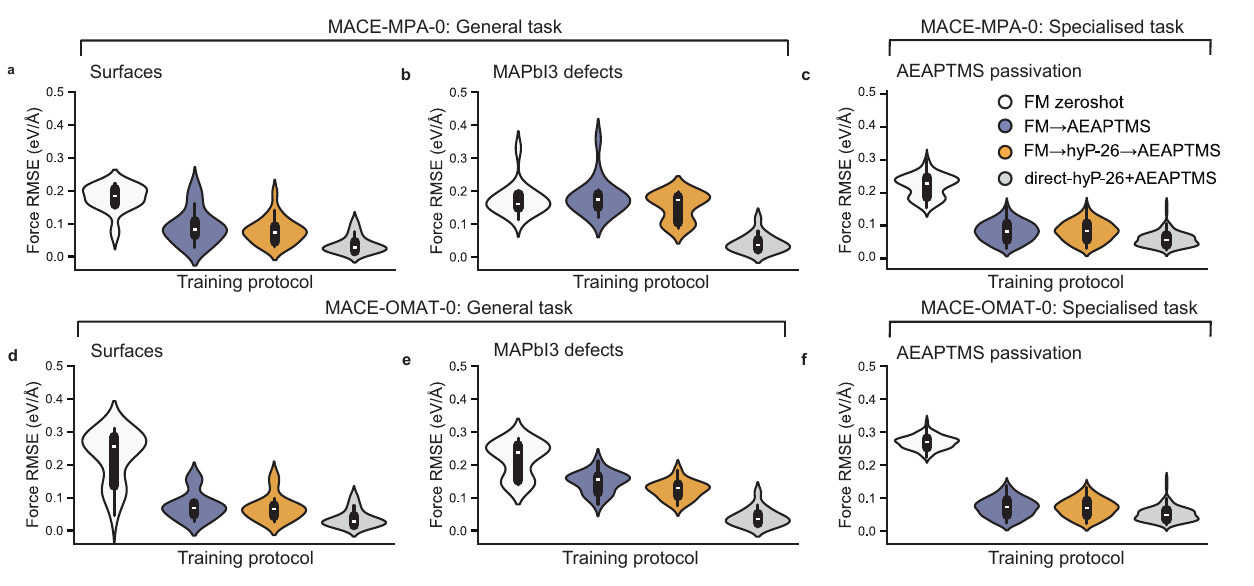}
\caption{Force component RMSE relative to DFT for MLIPs trained using different fitting approaches. Violin plots show errors for surface structures (`hyP-26' test set), \ce{MAPbI3} defect structures (`hyP-26' test set), and the AEAPTMS (\textbf{D2}) specialised test set. Panels (\textbf{a})--(\textbf{c}) use \texttt{MACE-MPA-0} as the starting FM, while panels (\textbf{d})--(\textbf{f}) use \texttt{MACE-OMAT-0}.}
  \label{fgr:SI-RMSE}
\end{figure}

\begin{figure}[H]
\centering
  \includegraphics[width=\textwidth]{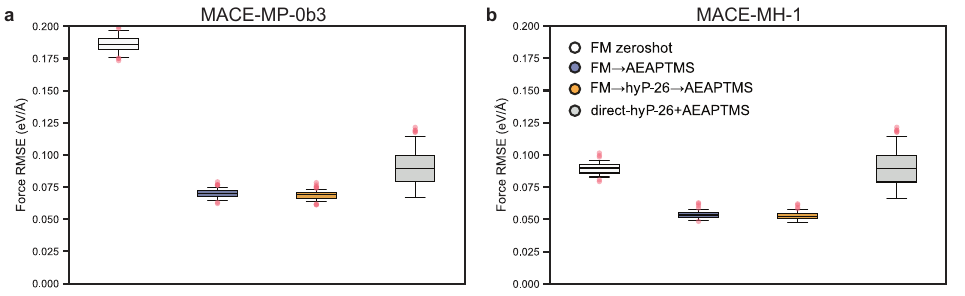}
  \caption{Root mean square error (RMSE) of force components predicted using the MLIPs obtained with different potential fitting approaches compared to DFT. (\textbf{a}), (\textbf{b}) Box plots showing RMSE on a MAPTMS specialised dataset, using \texttt{MACE-MP-0b3} and \texttt{MACE-MH-1} as the starting FMs, respectively. Outliers, representing data that fall outside of the upper and lower bounds of the boxplots, are shown in orange.}
  \label{fgr:RMSE-MAPTMS}
\end{figure}

\begin{figure}[H]
\centering
  \includegraphics[width=\textwidth]{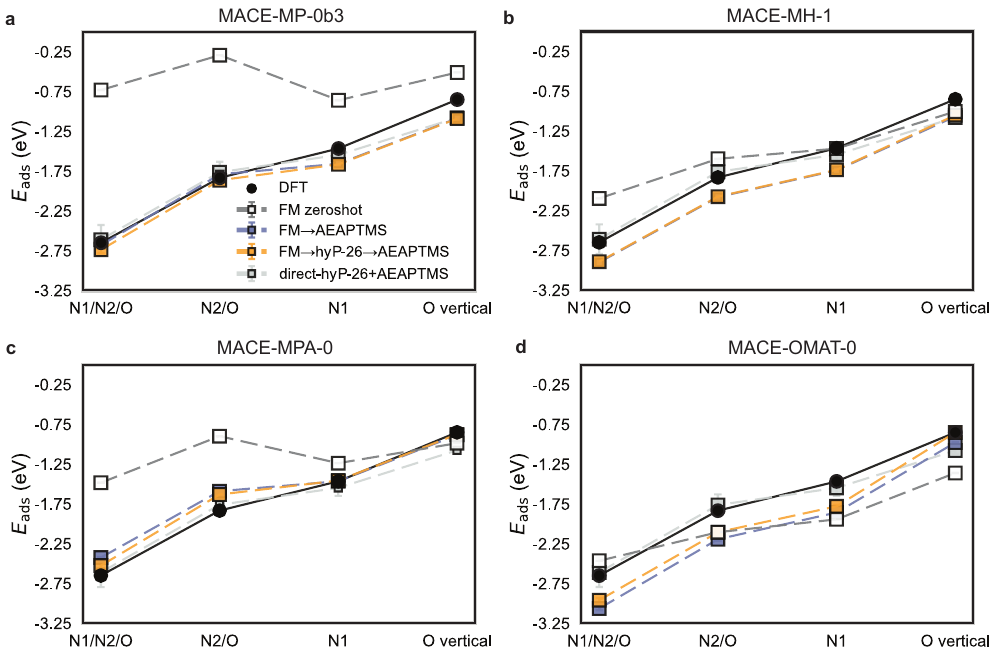}
  \caption{The MLIPs' performance in the prediction of $E_{\textnormal{ads}}$ evaluated on DFT-relaxed structures. (\textbf{a}) MLIP adsorption energy predictions compared to the DFT energy predictions for the DFT-relaxed systems with coordination types `N1/N2/O', `N2/O', `N1' and `O vertical', using \texttt{MACE-MP-0b3} as the starting FM. (\textbf{b}) Same, but using \texttt{MACE-MH-1} as the starting FM. (\textbf{c}) Same, but using \texttt{MACE-MPA-0} as the starting FM. (\textbf{d}) Same, but using \texttt{MACE-OMAT-0} as the starting FM.}
  \label{fgr:E_ads_singlepoint_FMs}
\end{figure} 

\begin{figure}[H]
\centering
  \includegraphics[width=\textwidth]{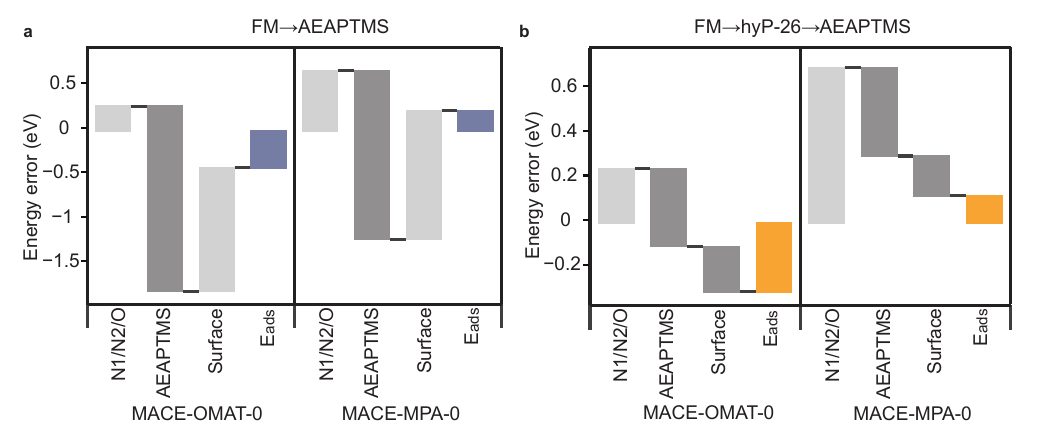}
\caption{Waterfall plots breaking down the energy errors contributing to the MLIPs' deviation from DFT-predicted $E_{\textnormal{ads}}$ for the `N1/N2/O' system. (\textbf{a}) Energy errors of the `FM$\rightarrow$AEAPTMS' fine-tuning protocol when using \texttt{MACE-OMAT-0} and \texttt{MACE-MPA-0} as starting FMs. (\textbf{b}) Same, but for the `FM$\rightarrow$hyP-26$\rightarrow$AEAPTMS' fine-tuning protocol.}
  \label{fgr:SI-waterfall-plots}
\end{figure}

\begin{figure}[H]
\centering
  \includegraphics[width=\textwidth]{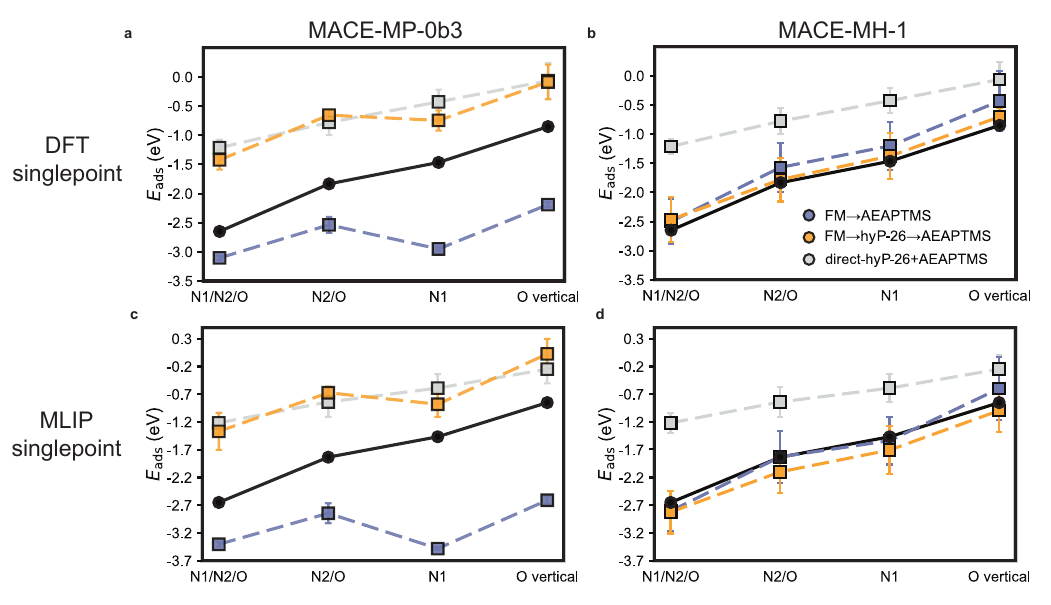}
  \caption{The MLIPs' performance in the relaxation of the passivated system in the expected coordination types. (\textbf{a}), (\textbf{b}) The DFT predictions of  $E_{\textnormal{ads}}$ for the MLIP-relaxed structures using \texttt{MACE-MP-0b3} and \texttt{MACE-MH-1} as the starting FMs, respectively. (\textbf{c}), (\textbf{d}) The MLIP predictions of  $E_{\textnormal{ads}}$ for the MLIP-relaxed structures using \texttt{MACE-MP-0b3} and \texttt{MACE-MH-1} as the starting FMs, respectively.  }
  \label{fgr:relaxation-singlepoint-DFT}
\end{figure}

\begin{figure}[H]
\centering
  \includegraphics[width=\textwidth]{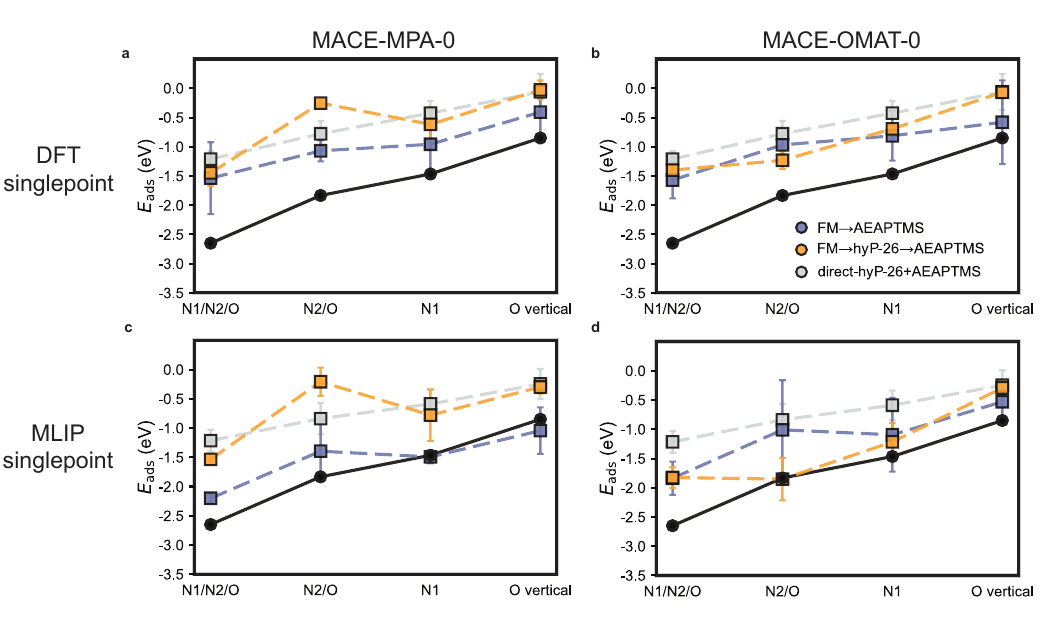}
  \caption{The MLIPs' performance in the relaxation of the passivated system in the expected coordination types. Panels (\textbf{a}), (\textbf{b}) show the DFT predictions of  $E_{\textnormal{ads}}$ of the MLIP-relaxed structures using \texttt{MACE-MPA-0} and \texttt{MACE-OMAT-0} as the starting FMs, respectively. Panels (\textbf{c}), (\textbf{d}) show the MLIP predictions of  $E_{\textnormal{ads}}$ of the MLIP-relaxed structures using \texttt{MACE-MPA-0} and \texttt{MACE-OMAT-0} as the starting FMs, respectively.  }
  \label{fgr:relaxation-singlepoint-DFT-mpa-omat}
\end{figure} 

\begin{figure}[H]
\centering
  \includegraphics[width=\textwidth]{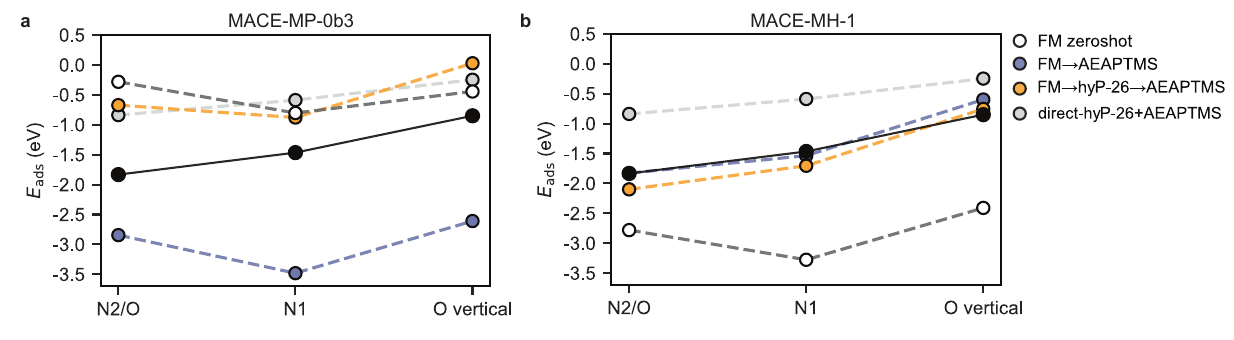}
  \caption{The MLIPs' performance in the relaxation of the passivated system in the expected coordination types, including the zero-shot benchmarks. (\textbf{a}) The DFT predictions of  $E_{\textnormal{ads}}$ of the MLIP-relaxed structures using \texttt{MACE-MP-0b3} as the starting FM. (\textbf{b}) Same, but using \texttt{MACE-MH-1} as the starting FM. The zero-shot models are not able to relax the system into the `N1/N2/O' coordination type, hence this structure is not shown here.}
  \label{fgr:relaxation-zeroshot}
\end{figure}

\begin{figure}[H]
\centering
  \includegraphics[width=\textwidth]{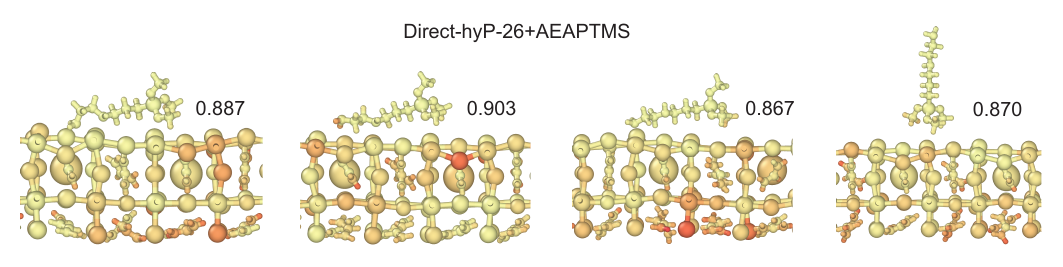}
  \caption{ Structural
similarity of the MLIP-relaxed structures, with relaxations driven by `direct-hyP-26+AEAPTMS', compared to the DFT-relaxed systems for each coordination type. Atom colors indicate their SOAP similarity to the corresponding atoms in the ground-truth DFT-
relaxed configuration (0: completely dissimilar; 1: identical), as visualized using OVITO.\cite{ovito}}
  \label{fgr:AEAPTMS-SOAP-direct}
\end{figure} 

\begin{figure}[H]
\centering
  \includegraphics[width=\textwidth]{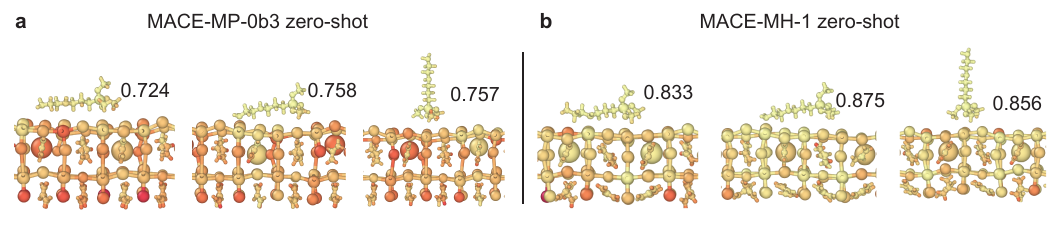}
  \caption{The MLIPs' performance in the relaxation of the passivated system in the expected coordination types. (\textbf{a}) Structural
similarity of the zero-shot \texttt{MACE-MP-0b3}-relaxed structures compared to the DFT-relaxed systems for each coordination
type. (\textbf{b}) Same, but for the zero-shot \texttt{MACE-MH-1} FM. Atom colors indicate their SOAP similarity to the corresponding atoms in the ground-truth DFT-
relaxed configuration (0: completely dissimilar; 1: identical), as visualized using OVITO.\cite{ovito} The zero-shot models are not able to relax the system into the `N1/N2/O' coordination type, hence this structure is not shown here.}
  \label{fgr:AEAPTMS-SOAP-zeorshot}
\end{figure}

\begin{figure}[H]
\centering
  \includegraphics[width=0.5\textwidth]{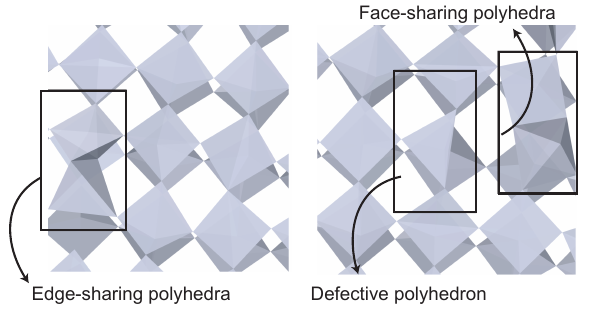}
  \caption{Edge-sharing polyehdra in simulations with passivation coverage of 1.28 molec/nm$^2$, as visualized using OVITO.\cite{ovito}}
  \label{fgr:defective-polyhedra}
\end{figure} 

\begin{figure}[H]
\centering
  \includegraphics[width=0.75\textwidth]{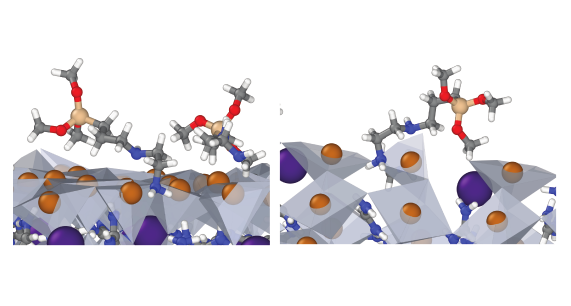}
  \caption{Simulation snapshots showing subsurface penetration of AEAPTMS molecules, as visualized using OVITO.\cite{ovito} The snapshots are extracted from simulations with passivation coverage values over 1 molec/nm$^2$.}
  \label{fgr:subsurface}
\end{figure}

\begin{figure}[H]
\centering
  \includegraphics[width=0.75\textwidth]{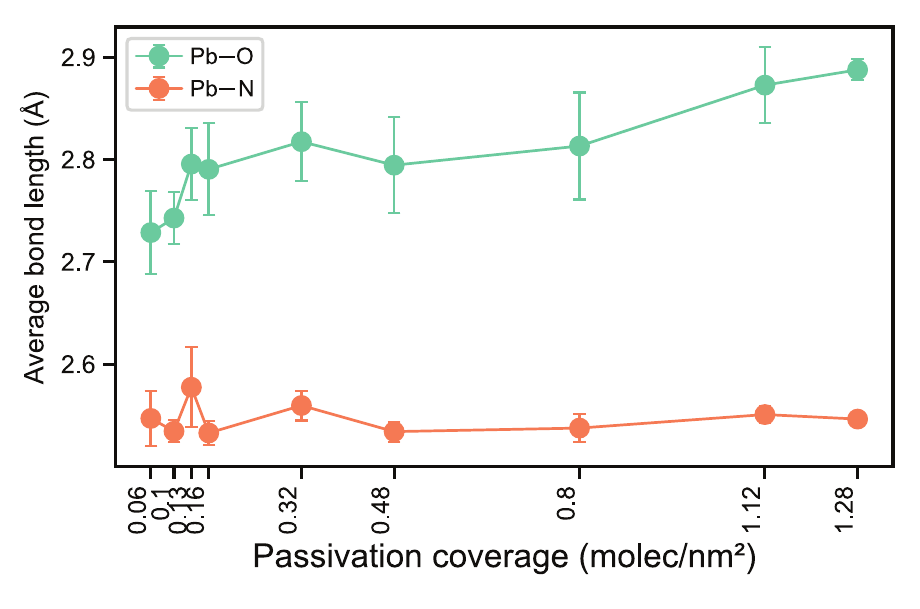}
  \caption{Influence of passivation coverage on average N--Pb and O--Pb bond-length, with the AEAPTMS molecule binding to the undercoordinated Pb at the perovskite surface.The intermolecular distances are averaged over all contacts up to a cut-off of 3.2~\AA{} for Pb--O and 3.0~\AA{} for Pb--N.}
  \label{fgr:bond-length}
\end{figure}